\documentclass[twoside]{IEEEtran}

\usepackage{cite}

\usepackage{amsmath,amssymb,amsfonts,mathrsfs,bm}

\usepackage{graphicx}
\usepackage{subfigure}
\graphicspath{{figures/}}

\usepackage{tabularx}
\usepackage{booktabs}
\usepackage{multicol}
\usepackage{multirow}

\usepackage{verbatim}

\makeatletter
  \newif\if@restonecol
  \def\@IEEEBIOskipN{5mm}
\makeatother

\usepackage[linesnumbered,ruled,lined]{algorithm2e}
\usepackage{algpseudocode}

\usepackage{scalerel}
\usepackage{tikz}
\usetikzlibrary{svg.path}

\definecolor{orcidlogocol}{HTML}{A6CE39}
\tikzset{
  orcidlogo/.pic={
    \fill[orcidlogocol] svg{M256,128c0,70.7-57.3,128-128,128C57.3,256,0,198.7,0,128C0,57.3,57.3,0,128,0C198.7,0,256,57.3,256,128z};
    \fill[white] svg{M86.3,186.2H70.9V79.1h15.4v48.4V186.2z}
                 svg{M108.9,79.1h41.6c39.6,0,57,28.3,57,53.6c0,27.5-21.5,53.6-56.8,53.6h-41.8V79.1z M124.3,172.4h24.5c34.9,0,42.9-26.5,42.9-39.7c0-21.5-13.7-39.7-43.7-39.7h-23.7V172.4z}
                 svg{M88.7,56.8c0,5.5-4.5,10.1-10.1,10.1c-5.6,0-10.1-4.6-10.1-10.1c0-5.6,4.5-10.1,10.1-10.1C84.2,46.7,88.7,51.3,88.7,56.8z};
  }
}

\newcommand\orcidicon[1]{%
  \href{https://orcid.org/#1}{
    \mbox{\scalerel*{
      \begin{tikzpicture}[yscale=-1, transform shape]
        \pic{orcidlogo};
      \end{tikzpicture}
    }{|}%
    }%
  }%
}

\newlength{\subfiglen}

\usepackage[colorlinks,allcolors=black]{hyperref} 

\title{Intelligent Resilience Testing for Decision-Making Agents with Dual-Mode Surrogate Adaptation}

\author{Jingxuan Yang\textsuperscript{\orcidicon{0000-0001-9798-7347}}, Weichao Xu\textsuperscript{\orcidicon{0009-0007-4691-1618}}, Yuchen Shi\textsuperscript{\orcidicon{0009-0006-1021-8478}}, Yi Zhang\textsuperscript{\orcidicon{0000-0001-5526-866X}}, \IEEEmembership{Senior Member,~IEEE},\\ Shuo Feng\textsuperscript{\orcidicon{0000-0002-2117-4427}}, \IEEEmembership{Member,~IEEE}, and Huaxin Pei\textsuperscript{\orcidicon{0000-0003-4815-2778}}%
\thanks{This work was supported in part by the National Natural Science Foundation of China under Grant 62503259. \textit{(Jingxuan Yang and Weichao Xu contributed equally to this work.) (Corresponding author: Huaxin Pei.)}}%
\thanks{Jingxuan Yang, Weichao Xu, Yuchen Shi and Huaxin Pei are with the Department of Automation, Tsinghua University, Beijing 100084, China (email: yangjx20@mails.tsinghua.edu.cn, xwc23@mails.tsinghua.edu.cn, shiyuche21@mails.tsinghua.edu.cn, phx17@tsinghua.org.cn).}%
\thanks{Yi Zhang is with the Department of Automation, Beijing National Research Center for Information Science and Technology (BNRist), Tsinghua University, Beijing 100084, China, and also with the Jiangsu Province Collaborative Innovation Center of Modern Urban Traffic Technologies, Nanjing 210096, China (e-mail: zhyi@mail.tsinghua.edu.cn).}%
\thanks{Shuo Feng is with the Department of Automation, Beijing National Research Center for Information Science and Technology (BNRist), Tsinghua University, Beijing 100084, China (e-mail: fshuo@tsinghua.edu.cn).}
}

\IEEEpubid{0000-0000~\copyright~2025 IEEE}

\begin{document}

\maketitle

\begin{abstract}
  Testing and evaluating decision-making agents remains challenging due to unknown system architectures, limited access to internal states, and the vastness of high-dimensional scenario spaces. Existing testing approaches often rely on surrogate models of decision-making agents to generate large-scale scenario libraries; however, discrepancies between surrogate models and real decision-making agents significantly limit their generalizability and practical applicability. To address this challenge, this paper proposes intelligent resilience testing (IRTest), a unified online adaptive testing framework designed to rapidly adjust to diverse decision-making agents. IRTest initializes with an offline-trained surrogate prediction model and progressively reduces surrogate-to-real gap during testing through two complementary adaptation mechanisms: (i) online neural fine-tuning in data-rich regimes, and (ii) lightweight importance-sampling-based weighting correction in data-limited regimes. A Bayesian optimization strategy, equipped with bias-corrected acquisition functions, guides scenario generation to balance exploration and exploitation in complex testing spaces. Extensive experiments across varying levels of task complexity and system heterogeneity demonstrate that IRTest consistently improves failure-discovery efficiency, testing robustness, and cross-system generalizability. These results highlight the potential of IRTest as a practical solution for scalable, adaptive, and resilient testing of decision-making agents.
\end{abstract}

\def\abstractname{Note to Practitioners}
\begin{abstract}
  This work considers the critical problem of efficiently testing and validating complex decision-making agents under unknown system architectures and high-dimensional scenario spaces. Efficient testing of decision-making agents is essential for improving system reliability and ensuring safety in real-world deployments, especially in safety-critical applications. Traditional intelligent testing methods mainly rely on offline-trained surrogate models to generate testing scenarios, which often suffer from poor generalization when applied to real agents with different structures or behaviors. In this work, an intelligent resilience testing framework is proposed that enables online adaptive testing by progressively correcting the surrogate-to-real gap through real-time testing feedback. This work shows that the proposed framework can achieve higher failure-discovery efficiency, stronger testing robustness, and better cross-system generalizability than conventional surrogate-based testing methods. As a practical consequence of this research, the proposed IRTest framework can effectively adapt surrogate models under both data-rich and data-limited testing conditions, enabling scalable and reliable testing with limited real-world interactions. This approach can also be applied to a wide range of AI-enabled systems, including autonomous driving, robotics, intelligent transportation systems, and other safety-critical decision-making applications.
\end{abstract}

\begin{IEEEkeywords}
  Intelligent resilience testing, decision-making agents, surrogate model, Bayesian optimization, adaptive scenario generation.
\end{IEEEkeywords}

\section{Introduction}
\label{sec:introduction}

\IEEEPARstart{T}{esting} and evaluating the performance of decision-making agents remain highly challenging in both development and deployment phases. These challenges stem from several fundamental factors: the internal architectures of many decision-making agents are unknown; obtaining real-time operational states can be difficult or impossible; and the testing scenario space is high-dimensional and complex \cite{delgado2021fuzz,nan2025safe,hua2025multi}. A widely adopted practice is to test decision-making agents in naturalistic testing environments (NTEs), observe their operational behaviors, and perform statistical analysis over large-scale testing results \cite{kalra2016driving}. However, NTE-based testing often exhibits low efficiency, high cost, and insufficient coverage of critical failure-prone scenarios—issues that become especially severe in safety-critical domains \cite{liu2024curse}. With the rapid advancement of machine learning, intelligent testing technology, grounded in the concept of ``AI for AI'', has demonstrated the ability to identify capability boundaries of decision-making agents and significantly reduce testing costs \cite{yan2023learning,ren2025digital,sun2021corner,li2024few,bai2024accurately,he2024knowmoformer,li2021scegene,wang2021advsim,menzel2018scenarios,tian2018deeptest,rempe2022generating,li2016intelligence,li2018artificial,li2019parallel,li2020theoretical,zhao2023genetic,riedmaier2020survey,li2025test}. Many existing approaches rely on surrogate models of decision-making agents to generate large-scale testing scenario libraries \cite{zhao2016accelerated,zhao2017accelerated,feng2020parti,feng2020partii,feng2021intelligent,feng2023dense}. Such knowledge-based or AI-based surrogate models provide an essential mechanism for accelerating testing in high-dimensional spaces while reducing reliance on real-world testing resources.

\IEEEpubidadjcol

Despite these benefits, a key obstacle limits the practical deployment of intelligent testing technologies: offline-generated scenario libraries frequently fail to generalize when applied to decision-making agents whose architectures, algorithms, or behavioral characteristics differ from the surrogate models used during scenario library generation. This surrogate-to-real gap leads to a severe degradation in testing performance, hindering the applicability of intelligent testing to diverse decision-making agents. Consequently, a critical need has emerged for intelligent testing methods that can rapidly adapt to different decision-making agents and remain effective even when only limited real testing interactions are available. Recent research has explored adaptive and learning-based testing techniques, including reinforcement learning, adversarial optimization, and generative modeling (see Section~\ref{sec:related_work} for a detailed review). However, most existing methods suffer from several limitations: (1) they often rely heavily on large-scale training data or specialized surrogate models; (2) they often lack efficient mechanisms to correct for discrepancies between surrogate models and real decision-making agents in high-dimensional scenario spaces. The core challenge remains unresolved: how to leverage surrogate models for efficient testing while dynamically correcting their prediction gaps during testing, ensuring both reliability and adaptability.

To address these gaps, this paper introduces the concept of intelligent resilience testing. IRTest aims to develop online adaptive testing methodologies that can rapidly adjust to different decision-making agents and progressively enhance testing effectiveness through real-time feedback. The proposed framework begins with a surrogate prediction model (SPM) trained offline and continually updates or reweights it using online testing results, thus reducing the surrogate-to-real gap during testing. Depending on testing resource availability, IRTest supports two operational modes: a data-rich regime, where the SPM is fine-tuned online via neural approximation; and a data-limited regime, where lightweight mixture models with importance-sampling-based weighting correction provide efficient adaptation. A Bayesian-optimization-based acquisition strategy coordinates scenario selection, enabling IRTest to balance exploration of new failure-prone regions and exploitation of high-risk areas predicted by the corrected surrogate model. The overall design philosophy of the proposed IRTest framework is illustrated in Fig.~1.

The main contributions of this study are summarized as follows:
\begin{itemize}
    \item \textbf{Intelligent Resilience Testing Framework.}
    We introduce the concept of IRTest, a unified online adaptive testing paradigm that systematically reduces surrogate-to-real gap and enhances testing robustness across diverse decision-making agents.
    \item \textbf{Dual-Mode Surrogate Adaptation Strategies.}
    We develop two complementary adaptation mechanisms: (i) an online fine-tuning scheme for data-rich regimes using neural surrogate approximation; and (ii) an importance-sampling-based weighting correction approach for data-limited regimes, ensuring rapid and efficient adaptation.
    \item \textbf{Bayesian Scenario Acquisition Function.}
    We design a Bayesian-optimization-driven scenario selection strategy that incorporates corrected surrogate predictions, enabling IRTest to efficiently explore and exploit high-risk regions of high-dimensional testing spaces.
    \item \textbf{Comprehensive Evaluation Across Testing Regimes.}
    We construct extensive experiments demonstrating that IRTest consistently improves failure discovery efficiency, testing stability, and cross-system generalizability.
\end{itemize}

The rest of the paper is organized as follows. Section~\ref{sec:related_work} reviews related literature and outlines key challenges in intelligent resilience testing. Section~\ref{sec:methods} formulates the problem and presents the proposed methodology. Section~\ref{sec:results} demonstrates the performance of the proposed framework across various testing conditions. Finally, Section~\ref{sec:conclusion} concludes the paper and discusses future research directions.

\begin{figure}[!t]
  \centering
  \includegraphics[width=.99\linewidth]{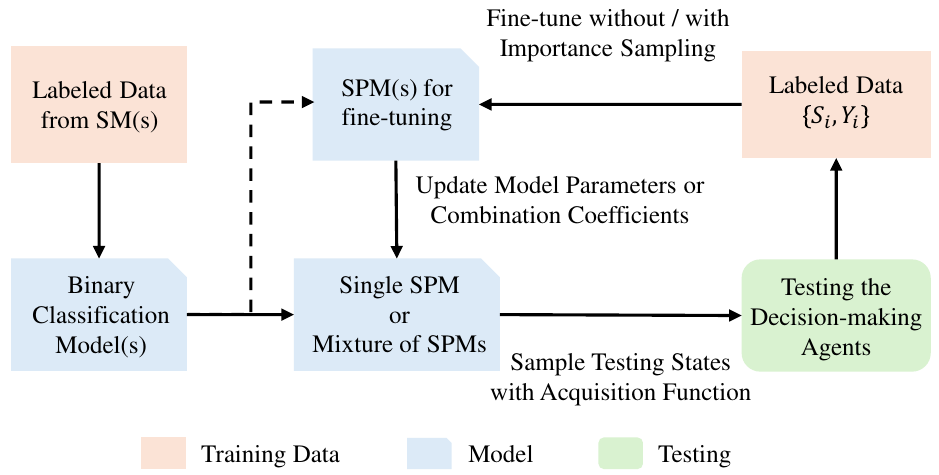}
  \caption{The workflow of IRTest.}
  \label{fig:simplified_method}
\end{figure}

\section{Related Work}
\label{sec:related_work}

In this section, we review related work from three perspectives: knowledge-based testing methods, AI-based testing methods, and adaptive testing methods.

\subsection{Knowledge-based Testing Methods}

Existing testing methods for decision-making agents often leverage domain knowledge or datasets to design specific scenario generation rules or frameworks.  A significant body of research focuses on establishing systematic architectures for testing scenario generation based on prior knowledge. For instance, \cite{kluck2018using,li2020ontology,duan2020test} employ ontologies to structurally represent scenarios, facilitating scenario generation through combinatorial techniques. While these methods offer simplicity and interpretability, they do not account for the dynamic interactions between intelligent agents within the environment.

Another category of approaches models the distributions of scenario parameters by analyzing existing datasets, enabling parameter sampling from these distributions \cite{sun2020scalability,yan2023learning,owen2013monte}. A commonly used technique in this context is Monte Carlo sampling, which selects parameter values without considering the actual characteristics of the system under test, thereby losing environment-specific contextualization.

Although knowledge-based testing methods perform well in automating testing scenario generation, they are inherently limited when applied to decision-making agents. Due to the high-dimensional nature of the testing space and the ``black-box'' characteristics of decision-making agents, knowledge-based methods often result in low testing efficiency and high resource costs, making it challenging to conduct comprehensive testing within limited resources.

\subsection{AI-based Testing Methods}

With the rapid advancement of artificial intelligence, AI-driven testing scenario generation has demonstrated significant advantages, particularly in accelerating the testing process and reducing costs. The main concept of AI-based testing is to learn to predict the criticality of testing scenarios, thereby efficiently generating critical testing scenarios while avoiding redundant testing. These methods are generally designed to achieve two primary testing objectives.

The first objective is to efficiently generate adversarial testing scenarios to expose capability deficiencies in the target system. Techniques such as gradient optimization, domain randomization, and generative models are employed to create adversarial image and point cloud data, aiding in the identification of weaknesses in system perception \cite{tu2020physically,abdelfattah2021towards,zhou2023corner}. Additionally, several studies \cite{ding2021multimodal,he2023adversarial,chen2021adversarial,khatiri2023simulation} formulate adversarial scenario generation as an optimization problem and leverage techniques such as deep reinforcement learning to evaluate the decision-making capabilities of decision-making agents.

Beyond adversarial testing, numerous methods have been proposed to accelerate the assessment of critical performance indicators, providing theoretical support for the practical deployment of decision-making agents \cite{feng2023dense,feng2021intelligent,feng2020parti,feng2020partii,zhao2016accelerated,zhao2017accelerated}. Approaches based on deep reinforcement learning and sparse adversarial sampling enable the efficient generation of critical testing scenarios in high-dimensional spaces while statistically ensuring the unbiasedness of test results. 

Compared to knowledge-based methods, AI-based testing offers significant advantages in improving testing efficiency and reducing costs. However, these methods heavily depend on the scale and quality of training data, limiting their adaptability to new decision-making agents.

\subsection{Adaptive Testing Methods}

To leverage testing information from other decision-making agents, increasing attention has been given to the generation of test scenarios using surrogate models. However, due to discrepancies between surrogate models and the actual systems under test, scenario libraries generated in this manner often lack generalizability, limiting their applicability to decision-making agents with different mechanisms and performance characteristics.

In this context, developing adaptive testing methods is crucial to enable the online optimization and adaptation of testing scenario libraries based on the specific characteristics of the system under test. Existing adaptive search algorithms—such as those based on Gaussian process regression \cite{feng2020adaptive,gong2023adaptive}, reinforcement learning \cite{koren2018adaptive}, unsupervised clustering \cite{mullins2018adaptive,mullins2017automated}, and gradient boosting trees \cite{sun2021adaptive,chen2016xgboost}—can generate challenging and diverse test scenarios online for certain decision-making agents without relying on surrogate models \cite{mullins2017delivering}. However, these methods are largely restricted to simple scenarios and struggle with inefficiencies in high-dimensional parameter spaces, making them insufficient for testing decision-making agents in complex applications. While some approaches \cite{yang2022adaptive,yang2023scv,yang2024denserl,yang2024adrl} have been designed to handle high-dimensional scenarios, their applicability remains largely confined to connected and automated vehicle systems.

In high-dimensional testing spaces, employing surrogate models to predict the criticality of testing scenarios becomes an essential strategy for intelligent testing. However, the inherent discrepancy between surrogate models and real-world systems remains a fundamental challenge, leading to poor generalizability in intelligent testing techniques. To address this, this paper introduces the concept of intelligent resilience testing, which aims to adaptively refine the testing scenario library generated from surrogate models, thereby enabling efficient and comprehensive testing of the system under test.

\section{Intelligent Resilience Testing}
\label{sec:methods}

\begin{figure*}[!t]
  \centering
  \includegraphics[width=.99\textwidth]{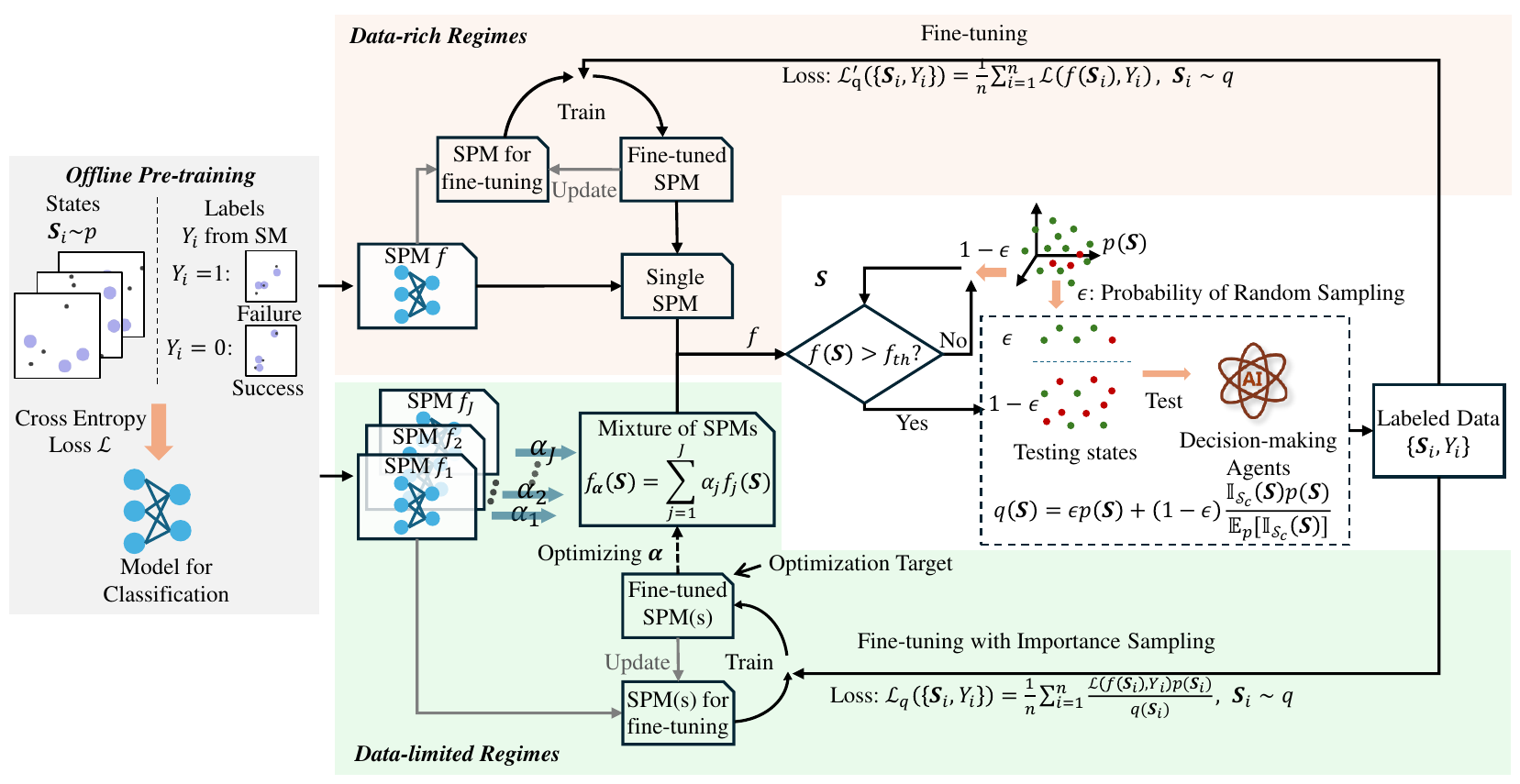}
  \caption{Illustration of the IRTest framework.}
  \label{fig:method}
\end{figure*}

In this section, we develop IRTest, an online adaptive testing framework that incrementally optimizes the SPMs with the goal of effective and efficient testing for decision-making agents. IRTest trains neural models as SPMs and continuously adapts them during testing, as illustrated in Fig.~\ref{fig:method}. Specifically, the problem formulation is presented in Section~\ref{subsec:problem_formulation}. Section~\ref{subsec:offline_pre_training} introduces the offline pre-training of the SPM on the surrogate decision-making agent. The adaptive testing procedures for data-rich regimes are detailed in Section~\ref{subsec:case1}, including the Bayesian-optimization-driven strategy for selecting the next testing state and the online fine-tuning of the SPM. Finally, Section~\ref{subsec:case2} addresses the challenges of data-limited regimes by presenting a combination-coefficient regression method enhanced with importance sampling.

\subsection{Problem Formulation}
\label{subsec:problem_formulation}

Let $f^*(\bm{s}):=\mathbb{P}_p(F|\bm{S}=\bm{s})\in[0,1]$, $\forall\bm{s}\in\mathcal{S}$ denote the probability of failure of the agents given the initial $m$ states $\bm{s}$, where $F$ denotes the failure event, $\mathcal{S}$ is the state space. We can use a surrogate prediction model $f$ to predict $f^*$. Define the surrogate-to-real gap as
\begin{equation}
  g(f^*||f)(\bm{s}):=f^*(\bm{s})-f(\bm{s}),~\forall\bm{s}\in\mathcal{S}.
\end{equation}

The goal of IRTest is to minimize the surrogate-to-real gap $g(f^*||f)$ and increase the testing efficiency. To achieve this goal, we use testing results to online optimize $f$.

\subsection{Offline Pre-training}
\label{subsec:offline_pre_training}

In IRTest, we employ an SPM $f$, which maps the testing state space $\mathcal{S}$ to failure probabilities in the range [0,1], to predict the ground truth $f^*$. The SPM is derived from a surrogate agent model (SAM) operating in the same application environment as the decision-making agent under test. Due to the complexity of the environment and the black-box nature of decision-making agents, we propose using a neural network (NN), such as a Multi-Layer Perceptron (MLP), as the SPM in IRTest. 

The task of predicting $f^*$ is modeled as a binary classification problem, where the two categories correspond to whether the agent under test fails in a given state. Using historical testing data from the SAM, we construct an offline training dataset by labeling each input state $\bm{s}$ with a binary label $Y$, where $Y=1$ indicates failure and $Y=0$ indicates success in scenarios initialized at state $\bm{s}$. By training the NN on this dataset, the network outputs a confidence score (failure probability) for each input $\bm{s}$, which is interpreted as the SPM, i.e., 
\begin{equation}
  f(\bm{s}) = \text{NN}(\bm{s})\in[0,1].
\end{equation}

Training the SPM $f$ is highly challenging due to the curse of rarity (CoR) \cite{liu2024curse}, i.e., the positive samples (failure cases) are so infrequent that an exceedingly large number of samples is required to obtain sufficient information about these rare events. Furthermore, training $f$ on a highly imbalanced dataset makes it difficult to achieve both high precision and high recall, especially when the imbalance ratio is large (e.g., greater than $10^6$). To mitigate the challenges posed by CoR, techniques such as class rebalancing can be applied, in which positive and negative samples are selected with a 1:1 ratio within each batch.

The pre-trained SPM is then used to test the target decision-making agent and is further optimized online during testing to reduce the surrogate-to-real gap and improve testing efficiency, as detailed in Subsections~\ref{subsec:case1} and \ref{subsec:case2}.

\subsection{Online Optimization in data-rich regimes}
\label{subsec:case1}

The failure probability predicted by the pre-trained SPM will be used to select testing states for evaluating the decision-making agent under test. Additionally, actual testing data from the target system will be leveraged to refine the SPM. In data-rich regimes, a large number of tests can be conducted on the target system to gather sufficient testing data. In this case, the workflow of IRTest will involve sampling testing states using a Bayesian-optimization-based strategy with bias-corrected acquisition functions and performing online fine-tuning based on the testing data, as shown in Fig.~\ref{fig:method}.

\subsubsection{Acquisition Function Design}
IRTest primarily focuses on states where $f(\bm{s})$ exceeds a given threshold, as these testing cases have a higher probability of triggering failures in the tested system, thereby increasing the proportion of critical testing scenarios among all sampled scenarios. 
The acquisition function is employed to select testing states based on prediction results. Specifically, denote $\mathcal{S}_c:=\{\bm{s}\in\mathcal{S}:f(\bm{s})>f_{\text{th}}\}$, where $f_{\text{th}}\in(0,1)$ is a threshold. The acquisition function is then given by
\begin{equation}
  \label{eq:acquisition_function}
  q(\bm{s})=\epsilon p(\bm{s}) + (1-\epsilon) \frac{\mathbb{I}_{\mathcal{S}_c}(\bm{s})p(\bm{s})}{\mathbb{E}_p[\mathbb{I}_{\mathcal{S}_c}(\bm{S})]},~\forall\bm{s}\in\mathcal{S},
\end{equation}
where $\epsilon\in(0,1)$ is a parameter that balances exploration and exploitation, and $p$ denotes the naturalistic distribution. To sample a testing state from $q$, we first sample a testing state $\bm{s}$ from $p$. Then, with probability $\epsilon$, we use $\bm{s}$ directly as the testing state. Otherwise, we continue sampling states $\bm{s}'$ from $p$ until $\bm{s}'\in\mathcal{S}_c$, at which point $\bm{s}'$ becomes the selected testing state. 

This sampling process, governed by the distribution $q$ and informed by predicted criticality, enables targeted exploration of the testing space, increasing the likelihood of triggering failure events. Furthermore, the introduction of the random sampling probability $\epsilon$ ensures that exploration is not overly constrained by focusing exclusively on $\mathcal{S}_c$. It is worth noting that this sampling distribution $q$ can guarantee 100\% coverage of the failure states with probability one, as it always samples testing states from $p$ with probability $\epsilon$.

\subsubsection{Fine-tuning of SPM}
To identify the failure cases of agents with high efficiency and high coverage, we propose generating testing states with $q$. The performance of the target system is then evaluated in scenarios initiated from these testing states.   
To enhance the precision and recall of the SPM for the decision-making agent under test, the SPM is fine-tuned by training on the set of visited states $\mathcal{D}$.

Each testing state $\bm{s}_i$ sampled with $q$, along with the binary label $Y_i$ determined by the actual performance of the agents under test, is collected as a pair of data to form $\mathcal{D}$. We train the SPM on $\mathcal{D}$ similar to the offline pre-training process. However, this fine-tuning is challenging due to the risk of catastrophic forgetting, where after being trained on $\mathcal{D}$, the SPM $f$ may forget the patterns learned before, leading to significant prediction errors. Existing methods designed to address catastrophic forgetting can be broadly categorized into three types: replay methods, regularization-based methods, and parameter-isolation methods \cite{de2021continual}. In this paper, we propose a simple yet effective approach to mitigate catastrophic forgetting. Specifically, we use a fixed number of training epochs to fine-tune the SPM at a fixed frequency.

The overall procedures of IRTest in data-rich regimes are summarized in Algorithm \ref{alg:IRTest_abundant}. During the testing process, each testing state is sampled from the distribution $q$, using the information of the surrogate model to enhance testing efficiency. The testing results are added to the set of visited states $\mathcal{D}$. Once a predefined number of tests $T_f$ have been conducted, the testing is paused and the SPM is trained on $\mathcal{D}$ for a fixed number of epochs $T_e$ to fine-tune it online. After online optimization, $\mathcal{S}_c$ is updated with the new $f$, and testing continues with new states generated from $q$. This process is repeated until the termination conditions defined by the tester are met.

\begin{algorithm}[!t]
  \label{alg:IRTest_abundant}
  \caption{Intelligent Resilience Testing for Data-Rich Regimes}
  \KwIn{naturalistic distribution $p$, SPM $f$, max simulation time $T$, max fine-tuning epoch $T_e$, fine-tuning frequency $T_f$, threshold $f_{\text{th}}\in(0,1)$, probability of random samplig $\epsilon$}
  \SetKwFunction{Uniform}{Uniform}
  \SetKwFunction{Sampling}{Sampling}
  \SetKwFunction{Testing}{Testing}
  \SetKwFunction{Train}{Train}
  
  Initialize the set of visited states $\mathcal{D} = \varnothing$\;
  $i\leftarrow0$\;
  \While{not termination}{
    $i\leftarrow i+1$\;
    \eIf{$\Uniform(0,1)<\epsilon$}
    {$\bm{s}_i \leftarrow \Sampling(p)$}
    {$\bm{s}_i \leftarrow \Sampling(p)$\;
    \While{$f(\bm{s}_i)<f_{\text{th}}$}{$\bm{s}_i \leftarrow \Sampling(p)$\;}
    }
    \tcp{Get label $Y_i$ by performing test with state $s_i$ and max simulation time $T$ }
    $Y_i \leftarrow \Testing(\bm{s}_i,T)$\;
    $\mathcal{D} \leftarrow \mathcal{D} \cup \{\bm{s}_i, Y_i\}$\;
    \uIf{$i \% T_f == 0$}
    {   
        \tcp{Train $f$ on $\mathcal{D}$ for $T_e$ epochs }
        $e \leftarrow0$\;
        \While{$e < T_e$}{
        $f \leftarrow \Train(f,\mathcal{D})$\;
        $e \leftarrow e+1$\;}
    }
    Update \textit{termination}\;
  }
\end{algorithm}

\subsection{Online Optimization in data-limited regimes}
\label{subsec:case2}

Due to the large number of parameters in neural networks, a substantial amount of data is required to achieve optimal results when fine-tuning the SPM directly on the training data. However, in practice, testing the target system often encounters the challenge of limited testing resources, primarily due to high testing cost, which necessitates using a small number of tests to achieve effective results. 
To address this challenge, an alternative weighting correction approach based on importance-sampling is proposed for data-limited regimes, as illustrated in Fig.~\ref{fig:method}. The weighting correction method shares the same acquisition function as described in Section \ref{subsec:case1} and follows similar testing and optimization procedures. The key distinction between online optimization methods in different regimes lies in the acquisition and optimization of $f$.

\subsubsection{Combination Coefficients Regression} \label{subsubsec:case2-1}
To reduce the testing data required by the IRTest process, we employ a mixture of SPMs to predict $f$. Specifically, the mixture of SPMs is given by 
\begin{equation}
  f_{\bm{\alpha}}(\bm{s})
  :=\sum_{j=1}^{J}\alpha_jf_j(\bm{s}),~\forall\bm{s}\in\mathcal{S},
\end{equation}
where $\bm{\alpha}=[\alpha_1,\dots,\alpha_J]^\top\in\mathbb{R}^J$ is the vector of combination coefficients, and $f_j$ represents the SPMs obtained from different SAMs. Thus, the prediction function $f$ in data-limited regimes can be viewed as the mixture of SPMs, i.e., $f(\bm{s}) = f_{\bm{\alpha}}(\bm{s})$. In this case, the online optimization problem reduces to the task of optimizing the combination coefficients $\bm{\alpha}$ of the SPMs. 

Let $\mathcal{D}$ represent the set of visited states. Each SPM $f_j$ is updated with newly acquired data $\mathcal{D}$ to obtain updated models, denoted as $\hat{f}_j$. These updated SPMs are then combined using the previously determined weights to optimize the combination coefficients. The corresponding optimization problem is formulated as:
\begin{equation}
  \label{eq:optimize_comb_coef_3}
  \begin{aligned}
    \min_{\bm{\alpha}}~&J_{\bm{\alpha}} := \frac{1}{2} \sum_{\bm{s} \in \mathcal{D}} \left[\sum_{j=1}^{J} \alpha_j^{\text{old}} \hat{f}_j(\bm{s}) - f_{\bm{\alpha}}(\bm{s})\right]^2 \\
    \text{s.t.}~~&\mathbf{1}^\top \bm{\alpha} = \mathbf{1}, \quad \bm{\alpha} \geqslant \mathbf{0}.
  \end{aligned}
\end{equation}

Note that this optimization problem is a quadratic programming (QP) problem whose optimal solutions can be found efficiently. By using combination coefficients regression, the number of parameters to be optimized is reduced to $J$, thereby lowering the data requirements. The key to solving these problems lies in training $\hat{f}_j$ with newly acquired data to obtain the optimal coefficients.

\subsubsection{Importance Sampling} \label{subsubsec:case2-2}

Due to the discrepancy between the naturalistic distribution $p$ and the sampling distribution $q$, the training data of the SPMs and the fine-tuning data $\mathcal{D}$ are not drawn from the same distribution. This mismatch introduces learning bias, as the models are trained on datasets with different sampling probabilities, potentially leading to suboptimal combination coefficients. To reduce the learning bias, we propose modifying the loss function by incorporating the importance sampling method \cite{owen2013monte}. Specifically, the original loss function is given by
\begin{equation}
  \mathcal{L}_p(\{\bm{S}_i,Y_i\})=\frac{1}{n}\sum_{i=1}^{n}\mathcal{L}(f(\bm{S}_i),Y_i),~\bm{S}_i\sim p,
\end{equation}
where $Y_i$ are the binary labels indicating whether the agent fails ($Y_i=1$) or succeeds ($Y_i=0$) in scenarios starting from state $\bm{S}_i$. Suppose that we use another sampling probability distribution $q$, the loss becomes
\begin{equation}
  \mathcal{L}_q'(\{\bm{S}_i,Y_i\})=\frac{1}{n}\sum_{i=1}^{n}\mathcal{L}(f(\bm{S}_i),Y_i),~\bm{S}_i\sim q.
\end{equation}
By applying importance sampling, the adjusted loss function is
\begin{equation}
  \mathcal{L}_q(\{\bm{S}_i,Y_i\})=\frac{1}{n}\sum_{i=1}^{n}\frac{\mathcal{L}(f(\bm{S}_i),Y_i)p(\bm{S}_i)}{q(\bm{S}_i)},~\bm{S}_i\sim q.
\end{equation}
According to importance sampling theory \cite{owen2013monte}, the importance sampling estimate is unbiased, i.e., $\mathbb{E}_p[\mathcal{L}_p(\{\bm{S}_i,Y_i\})]=\mathbb{E}_q[\mathcal{L}_q(\{\bm{S}_i,Y_i\})]$. Let $W:=p/q$ be the importance weight, which can be computed as
\begin{equation}
  \begin{aligned}
    W(\bm{s})
    &=\frac{p(\bm{s})}{q(\bm{s})}
    =\frac{p(\bm{s})}{\epsilon p(\bm{s}) + (1-\epsilon) \frac{\mathbb{I}_{\mathcal{S}_c}(\bm{s})p(\bm{s})}{\mathbb{E}_p[\mathbb{I}_{\mathcal{S}_c}(\bm{S})]}}\\
    &=\frac{1}{\epsilon + (1-\epsilon) \frac{\mathbb{I}_{\mathcal{S}_c}(\bm{s})}{\mathbb{E}_p[\mathbb{I}_{\mathcal{S}_c}(\bm{S})]}}\\
    &=\begin{cases}
      \dfrac{1}{\epsilon+\frac{1-\epsilon}{\mathbb{E}_p[\mathbb{I}_{\mathcal{S}_c}(\bm{S})]}},&\bm{s}\in\mathcal{S}_c,\\
      \dfrac{1}{\epsilon},&\bm{s}\in\mathcal{S}_{-c},
    \end{cases}
  \end{aligned}
\end{equation}
where $\mathcal{S}_{-c}=\mathcal{S}\setminus\mathcal{S}_{c}$. Furthermore, if we use the class rebalancing technique to train the SPM, then the original sampling probability distribution is changed to
\begin{equation}
  p'(\bm{s})=\begin{cases}
    \dfrac{p(\bm{s})}{2\mathbb{E}_p[\mathbb{I}_{\mathcal{S}_{+}}(\bm{S})]},&\bm{s}\in\mathcal{S}_{+},\\
    \dfrac{p(\bm{s})}{2\mathbb{E}_p[\mathbb{I}_{\mathcal{S}_{-}}(\bm{S})]},&\bm{s}\in\mathcal{S}_{-},
  \end{cases}
\end{equation}
where $\mathcal{S}_{+}$ is the set of positive samples and $\mathcal{S}_{-}$ is the set of negative samples. Similarly, the corresponding sampling distribution $ q' $ becomes
\begin{equation}
  q'(\bm{s})=\begin{cases}
    \dfrac{q(\bm{s})}{2\mathbb{E}_q[\mathbb{I}_{\mathcal{S}_{+}}(\bm{S})]},&\bm{s}\in\mathcal{S}_{+},\\
    \dfrac{q(\bm{s})}{2\mathbb{E}_q[\mathbb{I}_{\mathcal{S}_{-}}(\bm{S})]},&\bm{s}\in\mathcal{S}_{-},
  \end{cases}
\end{equation}
Therefore, the updated importance weight is
\begin{equation}
  \begin{aligned}
    W'(\bm{s})
    &=\frac{p'(\bm{s})}{q'(\bm{s})}\\
    &=\begin{cases}
      \dfrac{\mathbb{E}_q[\mathbb{I}_{\mathcal{S}_{+}}(\bm{S})]}{\left(\epsilon+\frac{1-\epsilon}{\mathbb{E}_p[\mathbb{I}_{\mathcal{S}_c}(\bm{S})]}\right)\mathbb{E}_p[\mathbb{I}_{\mathcal{S}_{+}}(\bm{S})]},&\bm{s}\in\mathcal{S}_c\cap\mathcal{S}_{+},\vspace{1em}\\
      \dfrac{\mathbb{E}_q[\mathbb{I}_{\mathcal{S}_{-}}(\bm{S})]}{\left(\epsilon+\frac{1-\epsilon}{\mathbb{E}_p[\mathbb{I}_{\mathcal{S}_c}(\bm{S})]}\right)\mathbb{E}_p[\mathbb{I}_{\mathcal{S}_{-}}(\bm{S})]},&\bm{s}\in\mathcal{S}_c\cap\mathcal{S}_{-},\vspace{1em}\\
      \dfrac{\mathbb{E}_q[\mathbb{I}_{\mathcal{S}_{+}}(\bm{S})]}{\epsilon\mathbb{E}_p[\mathbb{I}_{\mathcal{S}_{+}}(\bm{S})]},&\bm{s}\in\mathcal{S}_{-c}\cap\mathcal{S}_{+},\vspace{1em}\\
      \dfrac{\mathbb{E}_q[\mathbb{I}_{\mathcal{S}_{-}}(\bm{S})]}{\epsilon\mathbb{E}_p[\mathbb{I}_{\mathcal{S}_{-}}(\bm{S})]},&\bm{s}\in\mathcal{S}_{-c}\cap\mathcal{S}_{-}.\\
    \end{cases}
  \end{aligned}
\end{equation}

According to $W'(\bm{s})$, testing states belonging to different categories are assigned different importance weights. To compute these weights, three key values need to be numerically calculated: $\mathbb{E}_p[\mathbb{I}_{\mathcal{S}_{+}}(\bm{S})]$, $\mathbb{E}_q[\mathbb{I}_{\mathcal{S}_{+}}(\bm{S})]$ and $\mathbb{E}_p[\mathbb{I}_{\mathcal{S}_c}(\bm{S})]$. 
Here, $\mathbb{E}_p[\mathbb{I}_{\mathcal{S}_{+}}(\bm{S})]$ represents the failure rate when states are sampled according to the naturalistic distribution $p$, while $\mathbb{E}_q[\mathbb{I}_{\mathcal{S}_{+}}(\bm{S})]$ represents the failure rate when states are sampled according to the new sampling distribution $q$. The latter is related to the naturalistic failure rate and the precision of the mixture of SPMs $f$, i.e., $\mathbb{E}_q[\mathbb{I}_{\mathcal{S}_{+}}(\bm{S})] = \epsilon \mathbb{E}_p[\mathbb{I}_{\mathcal{S}_{+}}(\bm{S})] + (1-\epsilon) \mathrm{Precision}$. 
Finally, $\mathbb{E}_p[\mathbb{I}_{\mathcal{S}_c}(\bm{S})]$ represents the probability that the prediction of $f$ is positive when sampling states from $p$.

In practice, it is not possible to directly access the naturalistic distribution $p$ or the true precision of the SPM. To address this, we propose using the naturalistic distribution of the SAMs as an estimate for $p$, and rely on the testing data from the SPMs to compute the required values. Specifically, we assume that the testing data of the SPMs is sampled from $p$, and uses this data to estimate $\mathbb{E}_p[\mathbb{I}_{\mathcal{S}_{+}}(\bm{S})]$, $\mathbb{E}_q[\mathbb{I}_{\mathcal{S}_{+}}(\bm{S})]$ and $\mathbb{E}_p[\mathbb{I}_{\mathcal{S}_c}(\bm{S})]$. For instance, the probability of positive samples in the testing data is used as an estimate for the failure rate of the agent under test in states sampled from $p$. It is important to note that training data from the SPMs should not be used for this purpose, as the performance of the SPMs on the training data is typically overly optimistic, which deviates significantly from their performance in real-world scenarios.

Based on the combination of multiple SPMs and online optimization with importance sampling, the overall procedures of IRTest in Case II are summarized in Algorithm \ref{alg:IRTest_limited}.

\begin{algorithm}[!t]
  \label{alg:IRTest_limited}
  \caption{Intelligent Resilience Testing for Data-Limited Regimes}
  \KwIn{naturalistic distribution $p$, SPMs $f_j$ and their testing data $\mathcal{D}_j$, $j=1,\dots,J$, max simulation time $T$, max fine-tuning epoch $T_e$, fine-tuning frequency $T_f$, threshold $f_{\text{th}}\in(0,1)$, probability of random samplig $\epsilon$}
  \SetKwFunction{Uniform}{Uniform}
  \SetKwFunction{Sampling}{Sampling}
  \SetKwFunction{Testing}{Testing}
  \SetKwFunction{TrainIS}{Train\_with\_IS}

  Initialize the set of visited states $\mathcal{D} = \varnothing$ and combination coefficients $\bm{\alpha}=\left[\frac{1}{J}, \dots, \frac{1}{J}\right]$\;
  Estimate distribution $q$ with $\mathcal{D}_j$, $j=1,\dots,J$, see Section \ref{subsubsec:case2-2}\;

  $\hat{f}_j \leftarrow f_j$, $j=1,\dots,J$\;
  $i\leftarrow0$\;
  \While{not termination}{
    $i\leftarrow i+1$\;
    \eIf{$\Uniform(0,1)<\epsilon$}
    {$\bm{s}_i \leftarrow \Sampling(p)$}
    {$\bm{s}_i \leftarrow \Sampling(p)$\;
    \While{$\sum_{j=1}^{J}\alpha_jf_j(\bm{s}_i)<f_{\text{th}}$}{$\bm{s}_i \leftarrow \Sampling(p)$\;}
    }
    $Y_i \leftarrow \Testing(\bm{s}_i,T)$\;
    $\mathcal{D} \leftarrow \mathcal{D} \cup \{\bm{s}_i, Y_i\}$\;
    \uIf{$i \% T_f == 0$}
    {   
        \tcp{Train $\hat{f}_j$ with importance sampling on $\mathcal{D}$ for $T_e$ epochs, see Section \ref{subsubsec:case2-2} }
        $e \leftarrow0$\;
        \While{$e < T_e$}{

        $\hat{f}_j \leftarrow \TrainIS(\hat{f}_j,\mathcal{D})$, $j=1,\dots,J$\;

        $e\leftarrow e+1$\;}

        $f \leftarrow \sum_{j=1}^{J} \alpha_j \hat{f}_j$\;
        \tcp{ Define $f_{\bm{\alpha}} := \sum_{j=1}^{J}\alpha_jf_j$, and $J_{\bm{\alpha}} := \frac{1}{2} \sum_{\bm{s} \in \mathcal{D}} \left[f(\bm{s}) - f_{\bm{\alpha}}(\bm{s})\right]^2$, see Section \ref{subsubsec:case2-1}}
        $\bm{\alpha} \leftarrow \mathop{\arg\min}_{\bm{\alpha}:\mathbf{1}^\top \bm{\alpha} = \mathbf{1}, \bm{\alpha} \geqslant \mathbf{0}} \ J_{\bm{\alpha}}$
    }
    Update \textit{termination}\;
  }
\end{algorithm}

\section{Results}
\label{sec:results}

\subsection{Experimental Setup}

\begin{figure}[!t]
  \centering
  \setlength{\subfiglen}{4.3cm}
  \subfigure[PredatorPrey Environment]{\includegraphics[width=\subfiglen]{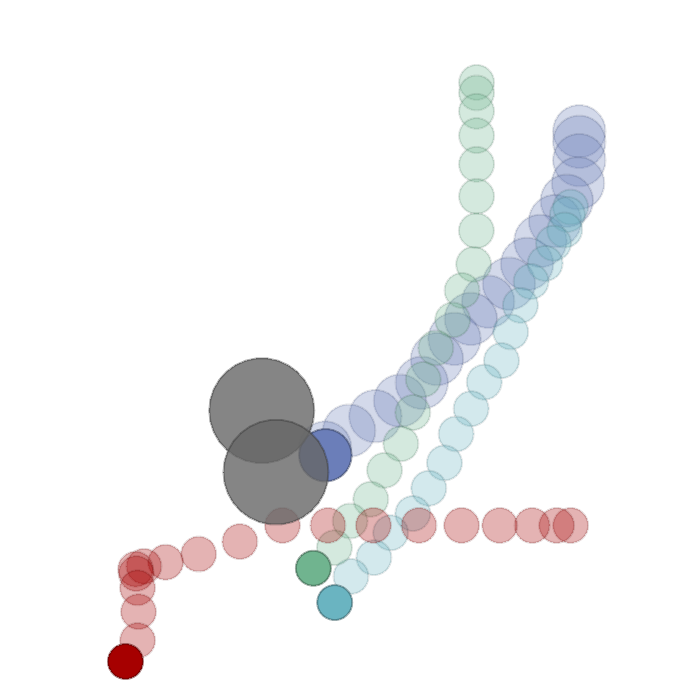}}
  \subfigure[CoopNavi Environment]{\includegraphics[width=\subfiglen]{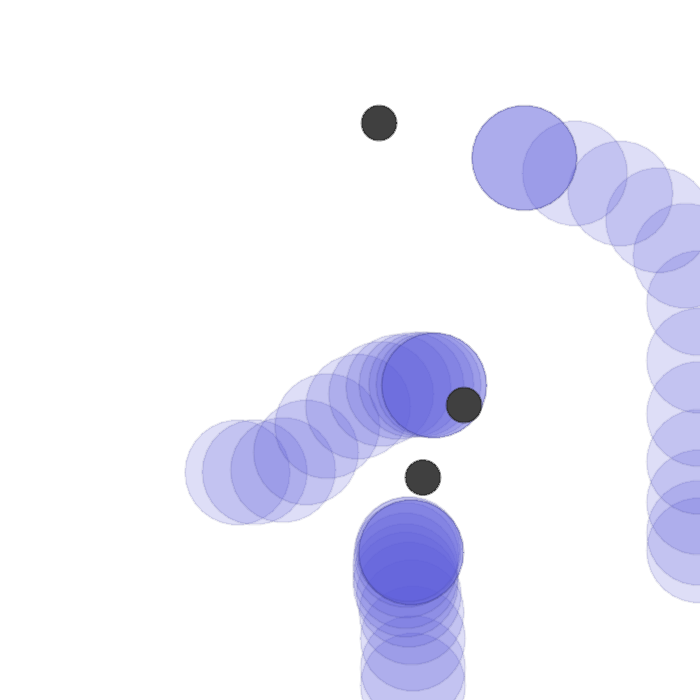}}
  \subfigure[BipedalWalker Environment]{\includegraphics[width=\subfiglen]{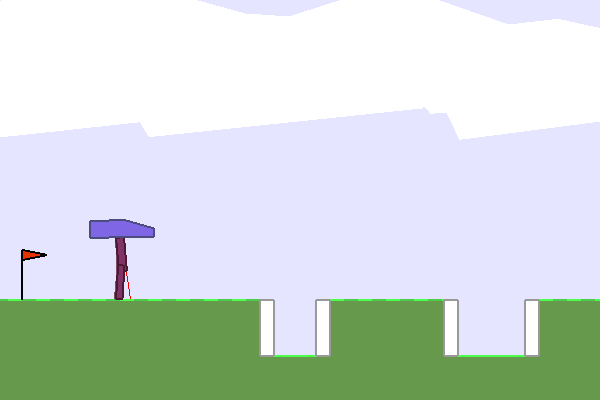}}
  \caption{Illustration of the PredatorPrey, CoopNavi, and BipedalWalker environments.}
  \label{fig:environment_visualizations}
\end{figure}

\subsubsection{Target Environments}
We conduct our IRTest in three well-established environments, as illustrated in Fig. \ref{fig:environment_visualizations}.

\textbf{PredatorPrey:} 
The PredatorPrey environment is a multi-agent game within the Multi-Agent Particle Environment (MPE) developed by OpenAI \cite{lowe2017multi}. In this setting, multiple predators must cooperate to capture prey on a map containing obstacles. Failure to capture the prey within a predefined time limit is considered a failure. In our setup, three predators are controlled by the target algorithm, and we introduce one prey and two obstacles. This configuration yields a 12-dimensional initial state, encompassing the positions of all six entities. To better capture the dynamic interactions between the target algorithm and the environment, we consider the first five frames of simulation collectively, concatenating them into a 60-dimensional testing state.

\textbf{CoopNavi:}
CoopNavi is another multi-agent environment within the MPE \cite{lowe2017multi}, where multiple agents collaborate to navigate toward designated landmarks while avoiding collisions. In our setup, the state at each frame is represented by a 12-dimensional vector, capturing the positions of three agents and three landmarks. Similar to the PredatorPrey environment, the testing state is constructed as a 60-dimensional vector by concatenating the states from the first five frames. A failure is recorded if the agents collide with one another or fail to reach the landmarks within the maximum allowed frames.

\textbf{BipedalWalker:} 
In the BipedalWalker environment \cite{brockman2016openai}, a bipedal agent must navigate diverse terrains while avoiding falls, which are considered failures. In our experiments, a fixed terrain sequence consisting of pits, stumps, and grass is used, with the length of each terrain segment treated as a parameter. This configuration produces a 10-dimensional testing state.

\begin{table*}[t]
  \centering
  \caption{The experimental details of different environments}
  \label{table:setup}
  \renewcommand{\arraystretch}{1.5}
  \begin{tabular}{|c|c|cccc|c|c|c|}
    \hline
    \multirow{2}{*}{\textbf{Environment}} & \multirow{2}{*}{$\bm{d}$} & \multicolumn{4}{c|}{\textbf{Parameters}}                                                                                                                & \multirow{2}{*}{\textbf{Training dataset (3:1:1)}} & \multirow{2}{*}{\textbf{Test dataset}} & \multirow{2}{*}{\textbf{SAM}} \\ \cline{3-6}
                                          &                           & \multicolumn{1}{c|}{$\bm{T}$} & \multicolumn{1}{c|}{$\bm{T_e}$}         & \multicolumn{1}{c|}{$\bm{T_f}$}            & \textbf{Termination}             &                                                    &                                           &                               \\ \hline
    \textbf{PredatorPrey}                 & 60                        & \multicolumn{1}{c|}{50}       & \multicolumn{1}{c|}{\multirow{3}{*}{2}} & \multicolumn{1}{c|}{\multirow{3}{*}{5000}} & \multirow{3}{*}{$1.5\times10^5$} & \multirow{2}{*}{$10^6$}                            & \multirow{3}{*}{$2\times10^5$}            & Different version             \\ \cline{1-3} \cline{9-9} 
    \textbf{CoopNavi}                     & 60                        & \multicolumn{1}{c|}{50}       & \multicolumn{1}{c|}{}                   & \multicolumn{1}{c|}{}                      &                                  &                                                    &                                           & Different version             \\ \cline{1-3} \cline{7-7} \cline{9-9} 
    \textbf{BipedalWalker}                & 10                        & \multicolumn{1}{c|}{300}      & \multicolumn{1}{c|}{}                   & \multicolumn{1}{c|}{}                      &                                  & $5\times10^5$                                      &                                           & Different algorithm           \\ \hline
    \end{tabular}
\end{table*}

\subsubsection{Experiment Design}
The high dimensionality and continuous nature of the testing states make exhaustive testing impractical. To address this issue, we construct a dataset by collecting a large number of testing states along with their corresponding failure labels. The dataset is then divided into training, validation, and test sets in a 3:1:1 ratio. We utilize this dataset to train a multi-layer perceptron model with the architecture MLP($d$, 256, 256, 256, 256, 1) for 100 epochs as the surrogate prediction model , where $d$ equals 60, 60, and 10 for PredatorPrey, CoopNavi, and BipedalWalker, respectively. The SPMs are trained using cross-entropy loss and the Adam optimizer. To mitigate the issue of label imbalance, we apply a class rebalancing technique, ensuring a 1:1 ratio of positive to negative samples in each batch. The failure labels in the training dataset for the SPM are obtained using a surrogate agent model, which is distinct from the agent under test. In PredatorPrey and CoopNavi, a different version of the agents is employed as the SAM, while in BipedalWalker, agents controlled by different algorithms are used as the SAMs. Experimental details are presented in Table~\ref{table:setup}.

\textbf{Data-rich regimes:}   
In data-rich regimes, we test the target agents following Algorithm \ref{alg:IRTest_abundant}. The maximum fine-tuning epoch is set to $T_e = 2$, the fine-tuning frequency to $T_f = 5000$, and testing terminates once the number of tests reaches a large value (e.g., 150,000). To objectively assess the effectiveness of IRTest, we pre-collect a dataset of 200,000 testing samples with real labels obtained from the tested agent. This dataset is used to evaluate the performance of the SPM after fine-tuning. 

\textbf{Data-limited regimes:} 
In this case, we employ a combination of three SPMs and optimize their combination coefficients according to Algorithm \ref{alg:IRTest_limited} to test the target agents under limited testing resources. Among the three SPMs, the SAMs for the first and third SPMs differ from the tested agent, while the SAM for the second SPM corresponds to the current agent. As a result, the optimal combination coefficient is expected to be [0, 1, 0]. As in data-rich regimes, we use the same parameters, $T_e$, $T_f$, and termination condition, and evaluate the system under test using a pre-collected test dataset. It is important to note that, in data-limited regimes, the primary focus is on the initial performance of IRTest. Setting the same maximum number of tests allows us to study the limiting performance of IRTest in this case. 

\subsection{Testing Results in Data-rich Regimes} \label{subsec:result1}

\begin{figure*}[!t]
  \centering
  \includegraphics[width=0.85\linewidth]{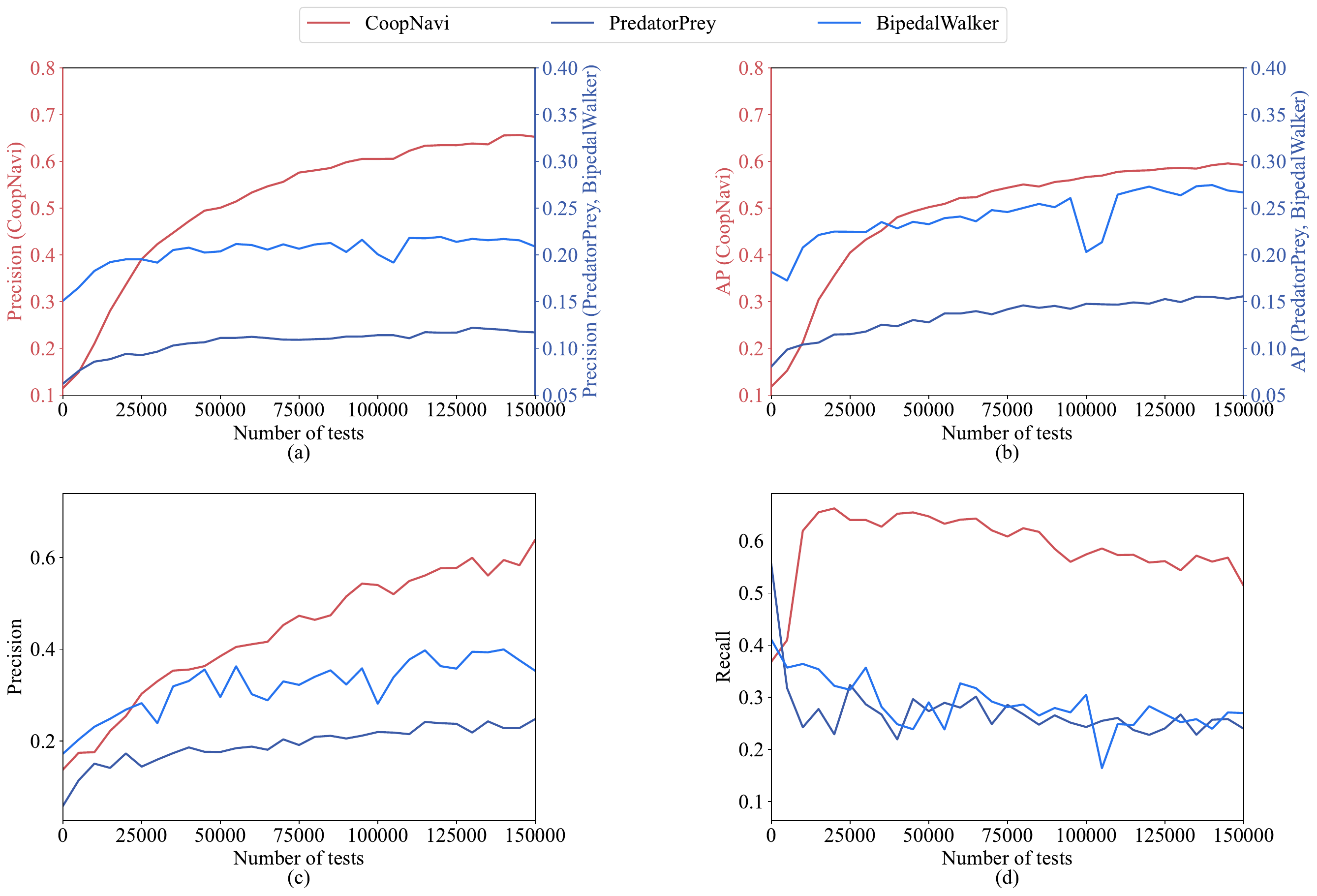}
  \caption{IRTest-R results: precision when recall is 0.5, average precision, precision and recall across the three environments, where $\epsilon=0.2$ and classification threshold $f_{\mathrm{th}} = 0.5$.}
  \label{fig:three_envs_abundant}
\end{figure*}

In this section, we evaluate the effectiveness of IRTest in data-rich regimes (denoted as IRTest-R) by analyzing the performance of the SPM on the test dataset as the number of tests increases. The evaluation considers four key aspects: (a) precision at a recall level of 0.5, (b) average precision (AP), (c) precision, and (d) recall at the current classification threshold. Metrics (a) and (b) are used to evaluate the overall performance of the SPM on the test dataset, while metrics (c) and (d) correspond to the operating points of the SPM, capturing its real-time performance during the IRTest process. Experimental results for a random probability of $\epsilon=0.2$ and a classification threshold of $f_{\mathrm{th}}=0.5$ across three environments are presented in Fig.~\ref{fig:three_envs_abundant}.

As illustrated in Fig.~\ref{fig:three_envs_abundant}(b), the AP of the SPM on the test dataset exhibits a steady improvement. Specifically, the AP in PredatorPrey, CoopNavi, and BipedalWalker increased from 0.0807 to 0.156, 0.119 to 0.592, and 0.182 to 0.267, corresponding to improvements of 93.3\%, 397\%, and 46.7\%, respectively. A higher AP signifies enhanced classification performance, achieving a more favorable trade-off between the failure rate of testing states and the coverage of the testing space. These results demonstrate that by adaptively fine-tuning the SPM during testing, IRTest effectively mitigates the surrogate-to-real gap.

\textbf{Selection of Hyper-parameters:} 
The random sampling probability $\epsilon$ and classification threshold $f_{\mathrm{th}}$ are crucial parameters for optimizing performance. Results for different values of $\epsilon$ and $f_{\mathrm{th}}$ are provided in Table \ref{table:abundant} in Appendix \ref{appendix:ablation_studies}. 

In IRTest, the random sampling probability is introduced to prevent the SPM from converging to a restricted region of the testing space,  thereby promoting a more thorough exploration. As shown in Table \ref{table:abundant}, the choice of $\epsilon$ significantly affects the fine-tuning of the SPM. Specifically, in the PredatorPrey environment, when $f_{\mathrm{th}}=0.5$, the AP values for $\epsilon = 0, 0.05, 0.2, 0.5$ eventually reach 0.127, 0.145, 0.156, and 0.149, respectively. Meanwhile, the precision at these operating points decreases from 0.347 to 0.154 as $\epsilon$ increases, while recall exhibits an increasing trend. The experimental results suggest that an excessively small $\epsilon$ leads to overreliance on the SPM’s predictions, restricting coverage to easily classified regions and thereby failing to adequately explore critical testing states. 
Conversely, a large $\epsilon$ diminishes the influence of the SPM, causing the sampling process to skew toward random sampling and thus reducing testing efficiency. 
Regarding the classification threshold $f_{\mathrm{th}}$, its impact on IRTest exhibits an inverse relationship to that of $\epsilon$. As shown in Table \ref{table:abundant}, a higher $f_{\mathrm{th}}$ combined with a lower $\epsilon$ leads to increased precision at the SPM operating point but at the expense of reduced recall.

The optimal parameter configuration across the three tested environments is $\epsilon=0.2$ and $f_{\mathrm{th}}=0.5$, striking a balance between exploration and efficiency. Based on this analysis, we recommend selecting moderate parameter values for practical applications. While these values generally yield the best overall performance, adjustments may be necessary depending on specific objectives. For instance, if the goal is to identify more critical test cases, slightly decreasing $\epsilon$ or increasing $f_{\mathrm{th}}$ helps raise the failure rate of the system under test. However, it is important to note that excessively high precision settings may exclude a substantial number of test states where $f(\bm{s})<f_{\mathrm{th}}$, thereby prolonging the classification phase.

\subsection{Testing Results in Data-limited Regimes} \label{subsec:result2}

\begin{figure*}[!t]
    \centering
    \includegraphics[width=0.85\linewidth]{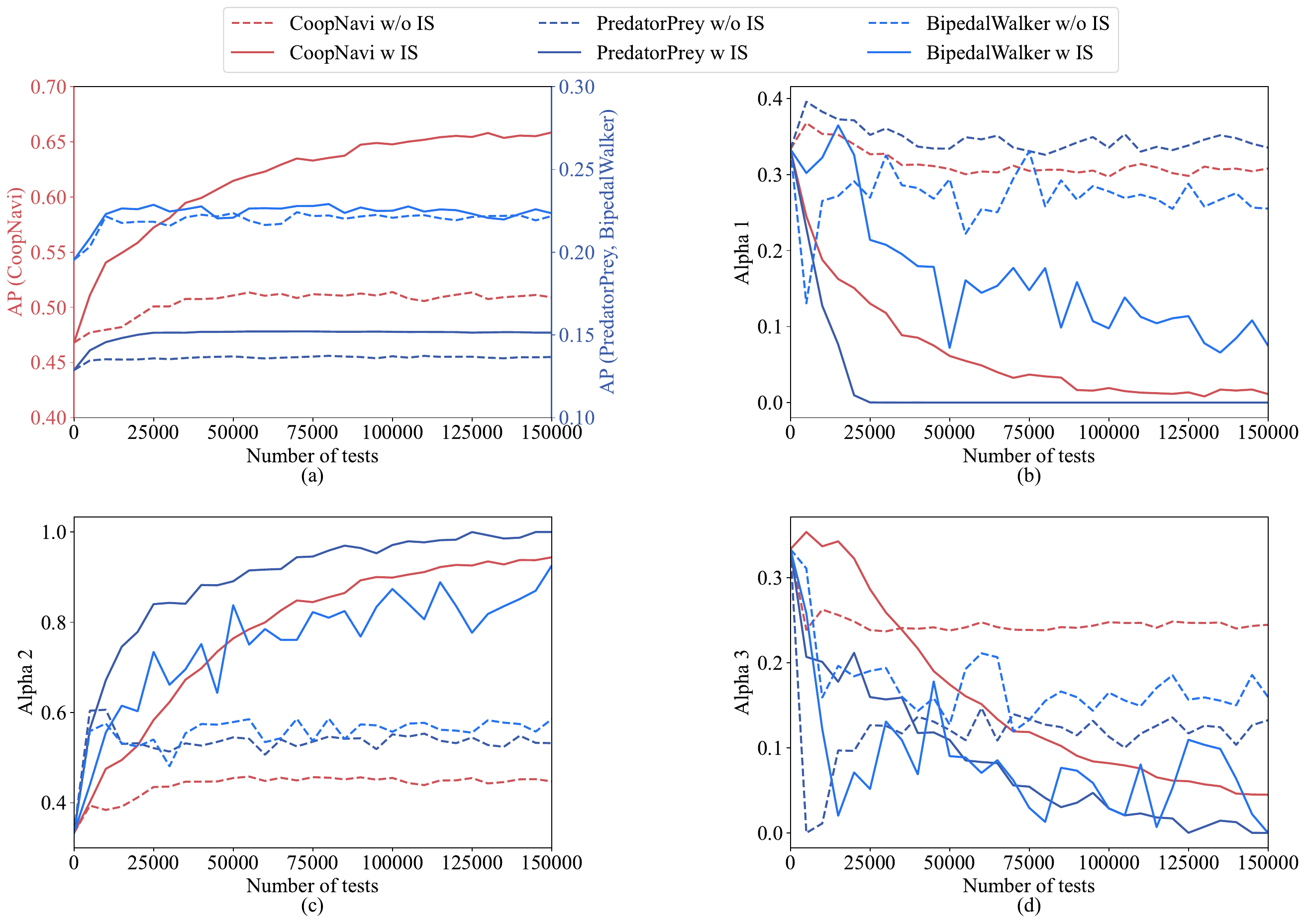}
    \caption{IRTest-L results: AP and combination coefficients across the three environments, where $\epsilon=0.05$ and classification threshold $f_{\mathrm{th}} = 0.5$.}
    \label{fig:three_envs_limited}
\end{figure*}

In scenarios with limited testing resources, IRTest employs a mixture of SPMs to predict $f$. To evaluate its effectiveness in this setting (denoted as IRTest-L), we analyze (a) the AP of the SPM mixture on the test dataset and (b-d) the combination coefficients $\alpha_1$, $\alpha_2$, and $\alpha_3$ for the three SPMs. Fig.~\ref{fig:three_envs_limited} illustrates the performance of IRTest-L across the three target environments. Furthermore, to assess the impact of importance sampling, comparison experiments with and without importance sampling are conducted, denoted as `w IS' and `w/o IS', respectively. 

Experimental results show that AP increases across all three environments as the number of tests grows, particularly in PredatorPrey and CoopNavi. Initially, AP rises sharply, reaching a high level quickly before stabilizing near its maximum value. This suggests that IRTest effectively reduces the surrogate-to-real gap with relatively few tests.  The results without importance sampling are generally lower than those with importance sampling, as importance sampling assigns varying weights based on sampling probability, thereby addressing issues related to training on datasets with differing sampling distributions.
The optimized combination coefficients for IRTest with importance sampling gradually converge toward the ground truth [0, 1, 0], reaching [0, 1, 1e-5], [0.011, 0.944, 0.045], and [0.075, 0.925, 1e-5], respectively. These findings indicate that IRTest effectively optimizes combination coefficients even under resource constraints, further demonstrating its efficacy.

To compare IRTest-L with IRTest-R, Table~\ref{table:ap} presents the AP of both methods using a limited number of tests (10k). Under resource constraints, IRTest-L achieves improvements of 40.4\%, 154\%, and 7.21\% over IRTest-R in the three environments, respectively. Table~\ref{table:ap} also reports the AP of the three SPMs after offline pre-training, where SPM-2, trained with data from the target agent, demonstrates the highest AP. By performing combination coefficient regression across all three SPMs, IRTest-L rapidly approximates the AP of SPM-2 with 10k tests and even surpasses it in BipedalWalker by effectively integrating multiple SPMs. These results highlight that Algorithm~\ref{alg:IRTest_limited}, by optimizing fewer coefficients, efficiently combines multiple SPMs and prioritizes the most effective models, ultimately leading to enhanced performance.

\textbf{Selection of Hyper-parameters:} 
Comprehensive ablation studies for different $\epsilon$ values and classification thresholds $f_{\mathrm{th}}$ are provided in Appendix~\ref{appendix:ablation_studies}. The results indicate that $\epsilon=0.05$ and $f_{\mathrm{th}}=0.5$ yield the best performance across all three environments.
When employing the importance-sampling-based weighting correction method, it is essential to ensure that the sub-models either include at least one well-performing candidate or can be effectively combined to achieve a superior model. In practice, the decision to use this method should consider the deviation between all SPMs and the optimal model $f^*$. If necessary, sub-models may require fine-tuning before performing coefficient optimization to achieve better performance.

\begin{table}[t]
  \centering
  \caption{the AP of different algorithms under a limited number of tests}
  \label{table:ap}
  \setlength{\tabcolsep}{2pt}
  \renewcommand{\arraystretch}{1.5}
  \begin{tabular}{cccccc}
    \hline
    \textbf{Environment}   & \textbf{SPM-1} & \textbf{SPM-2} & \textbf{SPM-3} & \textbf{\begin{tabular}[c]{@{}c@{}}IRTest-R\\ (10000)\end{tabular}} & \textbf{\begin{tabular}[c]{@{}c@{}}IRTest-L\\  (10000)\end{tabular}} \\ \hline
    \textbf{PredatorPrey}  & 0.0807         & 0.151          & 0.0738         & 0.104                                                               & 0.146                                                               \\
    \textbf{CoopNavi}      & 0.119          & 0.668          & 0.306          & 0.213                                                               & 0.541                                                               \\
    \textbf{BipedalWalker} & 0.182          & 0.214          & 0.123          & 0.208                                                               & 0.223                                                               \\ \hline
  \end{tabular}
  \end{table}

\subsection{Discovery of Failure Cases}

Sections~\ref{subsec:result1} and~\ref{subsec:result2} analyzed the performance of SPMs on the test dataset during IRTest. This section examines IRTest's ability to identify failure cases in data-rich and data-limited regimes. The precision of SPMs is closely linked to the failure rate across all testing states. Using Eq. \ref{eq:acquisition_function} and the analysis in Section \ref{subsubsec:case2-2}, we compare the failure rates of random testing and IRTest across three environments, with fixed parameters ($\epsilon=0.05$, $f_{\mathrm{th}}=0.5$) and a test budget of 150k cases.

\begin{figure*}[!t]
  \centering
  \subfigure[]{\includegraphics[width=0.3\linewidth]{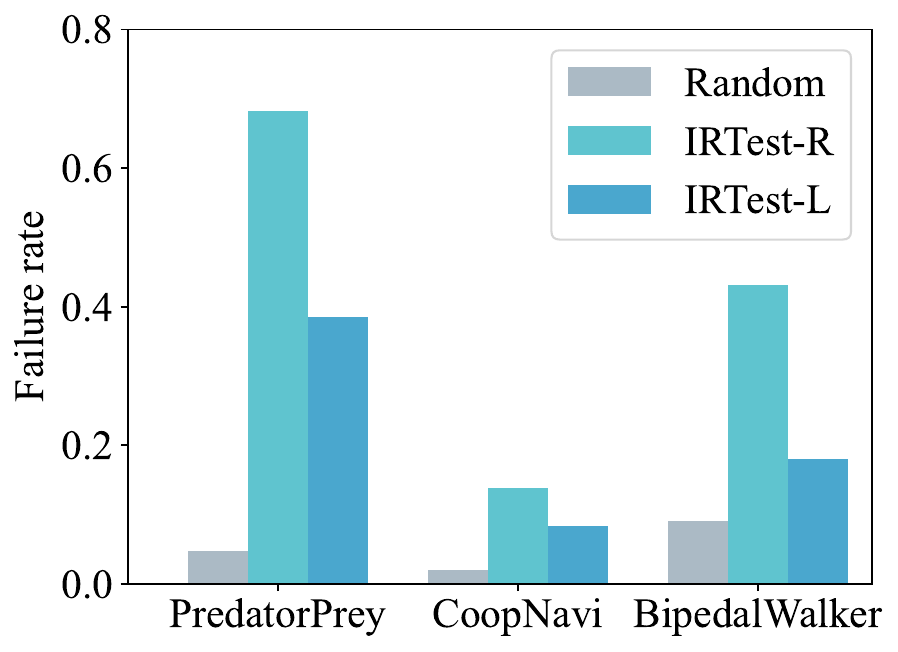}}
  \quad
  \subfigure[]{\includegraphics[width=0.3\linewidth]{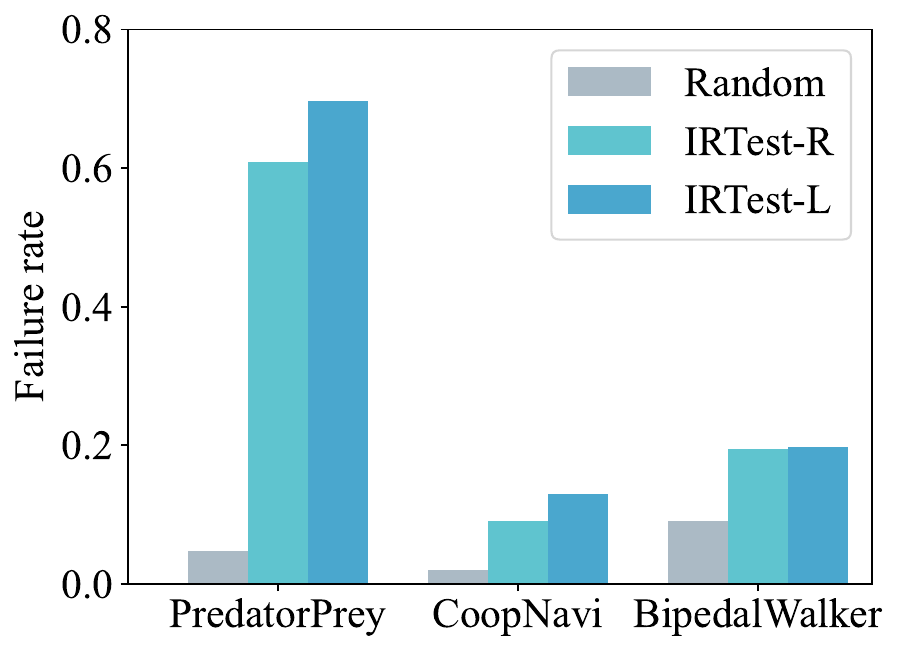}}
  \caption{Comparison of failure rates across different methods when (a) $f_{\mathrm{th}}=0.5$ or (b) recall = 0.5.}
  \label{fig:failure_rate}
\end{figure*}

\begin{figure*}[!t]
  \centering
  \subfigure[Failure case of PredatorPrey.]{\includegraphics[width=0.7\linewidth]{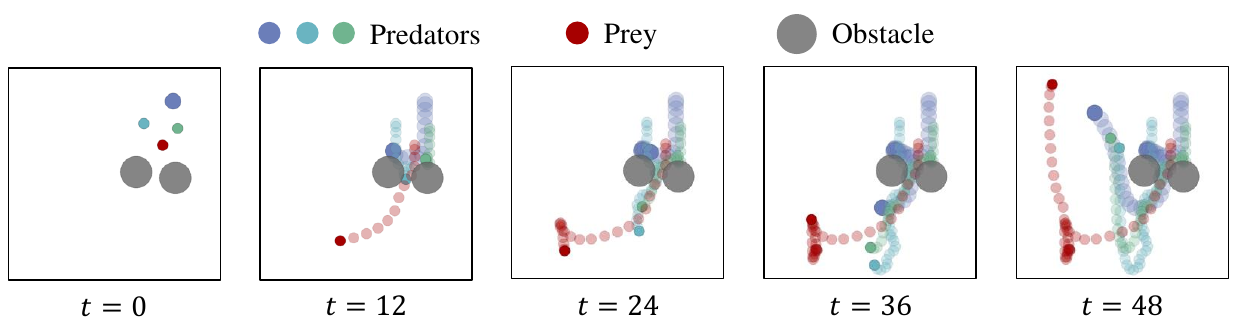}} 

  \subfigure[Failure case of CoopNavi.]{\includegraphics[width=0.7\linewidth]{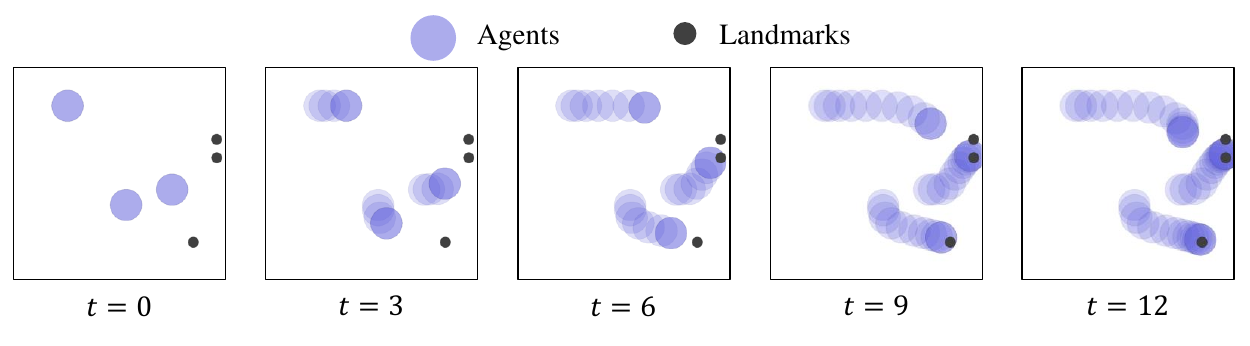}} 

  \subfigure[Failure case of BipedalWalker.]{\includegraphics[width=0.7\linewidth]{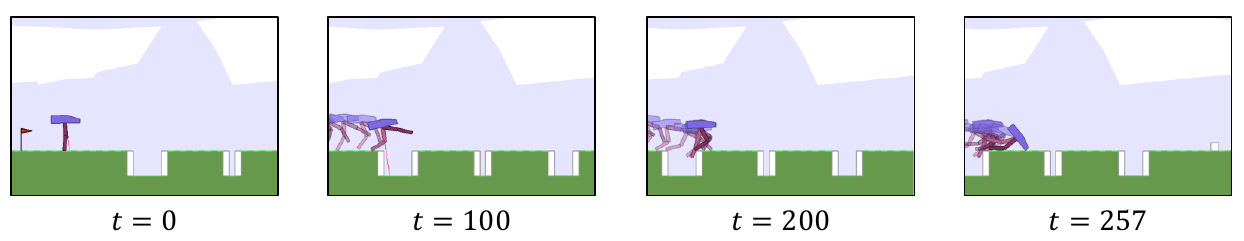}} 
  \caption{Examples of failure cases discovered by IRTest.}
  \label{fig:failure_examples}
\end{figure*}

Fig.~\ref{fig:failure_rate}(a) shows the failure rate at $f_{\mathrm{th}}=0.5$. Since IRTest maintains a fixed classification threshold during testing, these values represent actual failure rates. After fine-tuning, SPMs identify more critical testing states than random testing, leading to a higher failure rate. Additionally, IRTest-R achieves a higher failure rate than IRTest-L at $f_{\mathrm{th}}=0.5$, as its fine-tuned SPMs (based on Algorithm \ref{alg:IRTest_abundant}) continuously learn from new data with increasingly high failure rates, thereby improving precision step by step. In contrast, IRTest-L optimizes the combination coefficients of three SPMs instead of directly using the fine-tuned models. Additionally, importance sampling mitigates the impact of the new data distribution.  As shown in Fig.~\ref{fig:failure_rate}(b), when recall is fixed at 0.5, IRTest significantly outperforms random testing, and IRTest-L achieves the highest failure rate. These results demonstrate that IRTest can effectively enhance failure discovery in both cases when an appropriate classification threshold is chosen.  Examples of failure cases identified by IRTest are presented in Fig.~\ref{fig:failure_examples}.

It is important to note that IRTest-R prioritizes improving the precision of SPMs, which inevitably reduces recall. However, introducing $\epsilon$ ensures full test coverage given a sufficiently large number of test cases. Thus, in data-rich regimes, IRTest-R not only increases the failure rate but also ensures coverage of failure cases. By contrast, when the number of test cases is limited, recall becomes a more critical factor. In practical applications, selecting the appropriate algorithm and threshold depends on both the available test budget and the urgency of covering the test space.

\section{Conclusion and Future Work}
\label{sec:conclusion}

This paper introduced intelligent resilience testing, a unified framework designed to enhance the generalizability and adaptability of testing methods for decision-making agents. By integrating an offline surrogate prediction model with online adaptive correction mechanisms, IRTest effectively reduces surrogate-to-real gap during testing. Two complementary adaptation schemes were developed: a neural fine-tuning strategy for data-rich regimes and an importance-sampling-based weighting correction method for data-limited regimes. Furthermore, a Bayesian optimization strategy with bias-corrected acquisition functions enables efficient identification of failure-prone scenarios in high-dimensional testing spaces. Extensive experimental results demonstrate that IRTest significantly improves failure-discovery efficiency, testing stability, and robustness across heterogeneous decision-making agents.

Despite these promising results, several limitations warrant further research. First, the performance of IRTest still depends on the initial quality of the surrogate model; poor surrogate initialization may slow adaptation in highly complex systems. Second, while the Bayesian optimization module is effective, its scalability may diminish in extremely high-dimensional scenario spaces, motivating the development of more expressive acquisition functions or dimensionality-reduction-based kernels. Third, the current framework assumes consistent test outcome definitions across systems; extending IRTest to handle heterogeneous or multi-task performance metrics remains an important direction.

\appendix

\subsection{Ablation Studies} \label{appendix:ablation_studies}

Tables \ref{table:abundant}–\ref{table:limited} present the experimental results of IRTest-R and IRTest-A. As discussed in Section \ref{sec:results}, a moderate random sampling probability and classification threshold are required to achieve optimal AP improvement, though adjustments may be needed depending on specific objectives.

\begin{table*}[t]
  \centering
  \caption{Performance of IRTest-R with different $\epsilon$ and $f_\mathrm{th}$ across all the three environments}
  \label{table:abundant}
  \renewcommand{\arraystretch}{1.2}
  \begin{tabular}{c|c|c|ccc|ccc|ccc|ccc}
    \hline
    \multirow{2}{*}{\textbf{Environment}}   & \multirow{2}{*}{$\bm{\epsilon}$} & \multirow{2}{*}{$\bm{f_{\mathrm{th}}}$} & \multicolumn{3}{c|}{\textbf{Precision at recall=0.5}} & \multicolumn{3}{c|}{\textbf{AP}}           & \multicolumn{3}{c|}{\textbf{Precision}}    & \multicolumn{3}{c}{\textbf{Recall}}                            \\ \cline{4-15} 
                                            &                               &                               & \textbf{0k}    & \textbf{25k}    & \textbf{150k}   & \textbf{0k} & \textbf{25k} & \textbf{150k} & \textbf{0k} & \textbf{25k} & \textbf{150k} & \textbf{0k} & \textbf{25k} & \multicolumn{1}{c}{\textbf{150k}} \\ \hline
    \multirow{9}{*}{\textbf{PredatorPrey}}  & \textbf{0}                    & \textbf{0.5}                  & 0.0624         & 0.0847          & 0.0676          & 0.0807      & 0.115        & 0.127         & \textbf{0.0581}      & \textbf{0.230}        & \textbf{0.347}         & 0.555       & 0.135        & 0.0995                             \\ \cline{2-3}
                                            & \textbf{0.05}                 & \textbf{0.1}                  & 0.0624         & 0.0944          & 0.118           & 0.0807      & 0.112        & 0.133         & 0.0337      & 0.0320       & 0.0351        & \textbf{0.897}       & \textbf{0.922}        & 0.944                              \\ \cline{2-3}
                                            & \textbf{0.05}                 & \textbf{0.3}                  & 0.0624         & \textbf{0.106}           & \textbf{0.136}           & 0.0807      & 0.116        & 0.150         & 0.0443      & 0.0637       & 0.0796        & 0.734       & 0.714        & 0.705                              \\ \cline{2-3}
                                            & \textbf{0.05}                 & \textbf{0.5}                  & 0.0624         & 0.0881          & 0.0941          & 0.0807      & 0.114        & 0.145         & \textbf{0.0581}      & 0.186        & 0.315         & 0.555       & 0.216        & 0.144                              \\ \cline{2-3}
                                            & \textbf{0.2}                  & \textbf{0.5}                  & 0.0624         & 0.0929          & 0.117           & 0.0807      & 0.115        & \textbf{0.156}         & \textbf{0.0581}      & 0.144        & 0.248         & 0.555       & 0.324        & 0.240                              \\ \cline{2-3}
                                            & \textbf{0.5}                  & \textbf{0.1}                  & 0.0624         & 0.0942          & 0.113           & 0.0807      & 0.116        & 0.130         & 0.0337      & 0.0353       & 0.0325        & \textbf{0.897}       & 0.911        & \textbf{0.953}                              \\ \cline{2-3}
                                            & \textbf{0.5}                  & \textbf{0.3}                  & 0.0624         & 0.0960          & 0.121           & 0.0807      & 0.105        & 0.141         & 0.0443      & 0.0613       & 0.0681        & 0.734       & 0.704        & 0.756                              \\ \cline{2-3}
                                            & \textbf{0.5}                  & \textbf{0.5}                  & 0.0624         & 0.0913          & 0.128           & 0.0807      & 0.104        & 0.149         & \textbf{0.0581}      & 0.111        & 0.154         & 0.555       & 0.393        & 0.422                              \\ \cline{2-3}
                                            & \textbf{0.7}                  & \textbf{0.5}                  & 0.0624         & 0.0970          & 0.120           & 0.0807      & \textbf{0.117}        & 0.140         & \textbf{0.0581}      & 0.0994       & 0.114         & 0.555       & 0.487        & 0.520                              \\ \hline
    \multirow{10}{*}{\textbf{CoopNavi}}     & \textbf{0}                    & \textbf{0.5}                  & 0.115          & 0.392           & 0.657           & 0.119       & 0.400        & 0.580         & 0.137       & 0.344        & 0.766         & 0.368       & 0.568        & 0.396                              \\ \cline{2-3}
                                            & \textbf{0.05}                 & \textbf{0.1}                  & 0.115          & 0.374           & 0.628           & 0.119       & 0.388        & 0.584         & 0.109       & 0.0747       & 0.198         & \textbf{0.533}       & \textbf{0.943}        & 0.871                              \\ \cline{2-3}
                                            & \textbf{0.05}                 & \textbf{0.3}                  & 0.115          & 0.398           & \textbf{0.663}           & 0.119       & 0.410        & \textbf{0.599}         & 0.125       & 0.174        & 0.404         & 0.430       & 0.822        & 0.737                              \\ \cline{2-3}
                                            & \textbf{0.05}                 & \textbf{0.5}                  & 0.115          & 0.377           & 0.639           & 0.119       & 0.402        & 0.583         & 0.137       & 0.367        & 0.715         & 0.368       & 0.511        & 0.435                              \\ \cline{2-3}
                                            & \textbf{0.2}                  & \textbf{0.1}                  & 0.115          & 0.393           & 0.623           & 0.119       & 0.401        & 0.583         & 0.109       & 0.0880       & 0.174         & \textbf{0.533}       & 0.918        & \textbf{0.881}                              \\ \cline{2-3}
                                            & \textbf{0.2}                  & \textbf{0.3}                  & 0.115          & \textbf{0.411}           & 0.649           & 0.119       & \textbf{0.412}        & 0.594         & 0.125       & 0.169        & 0.338         & 0.430       & 0.826        & 0.777                              \\ \cline{2-3}
                                            & \textbf{0.2}                  & \textbf{0.5}                  & 0.115          & 0.391           & 0.653           & 0.119       & 0.405        & 0.592         & 0.137       & 0.303        & 0.639         & 0.368       & 0.640        & 0.514                              \\ \cline{2-3}
                                            & \textbf{0.2}                  & \textbf{0.7}                  & 0.115          & 0.323           & 0.454           & 0.119       & 0.368        & 0.493         & \textbf{0.148}       & \textbf{0.824}        & \textbf{0.936}         & 0.297       & 0.0536       & 0.141                              \\ \cline{2-3}
                                            & \textbf{0.5}                  & \textbf{0.5}                  & 0.115          & 0.396           & 0.636           & 0.119       & 0.401        & 0.587         & 0.137       & 0.265        & 0.469         & 0.368       & 0.705        & 0.671                              \\ \cline{2-3}
                                            & \textbf{0.7}                  & \textbf{0.5}                  & 0.115          & 0.370           & 0.616           & 0.119       & 0.395        & 0.575         & 0.137       & 0.225        & 0.426         & 0.368       & 0.757        & 0.701                              \\ \hline
    \multirow{7}{*}{\textbf{BipedalWalker}} & \textbf{0}                    & \textbf{0.5}                  & 0.151          & 0.197           & 0.181           & 0.182       & \textbf{0.233}        & 0.201         & \textbf{0.172}       & \textbf{0.374}        & 0.303         & 0.411       & 0.224        & 0.0447                             \\ \cline{2-3}
                                            & \textbf{0.05}                 & \textbf{0.5}                  & 0.151          & 0.188           & \textbf{0.222}           & 0.182       & 0.231        & 0.259         & \textbf{0.172}       & 0.352        & \textbf{0.451}         & 0.411       & 0.236        & 0.158                              \\ \cline{2-3}
                                            & \textbf{0.2}                  & \textbf{0.1}                  & 0.151          & 0.198           & 0.213           & 0.182       & 0.225        & 0.229         & 0.138       & 0.102        & 0.110         & \textbf{0.575}       & \textbf{0.950}        & \textbf{0.928}                              \\ \cline{2-3}
                                            & \textbf{0.2}                  & \textbf{0.3}                  & 0.151          & 0.200           & 0.205           & 0.182       & 0.226        & 0.226         & 0.156       & 0.144        & 0.148         & 0.478       & 0.717        & 0.727                              \\ \cline{2-3}
                                            & \textbf{0.2}                  & \textbf{0.5}                  & 0.151          & 0.195           & 0.209           & 0.182       & 0.225        & \textbf{0.267}         & \textbf{0.172}       & 0.282        & 0.353         & 0.411       & 0.315        & 0.270                              \\ \cline{2-3}
                                            & \textbf{0.5}                  & \textbf{0.5}                  & 0.151          & 0.203           & 0.208           & 0.182       & 0.225        & 0.245         & \textbf{0.172}       & 0.236        & 0.232         & 0.411       & 0.405        & 0.447                              \\ \cline{2-3}
                                            & \textbf{0.7}                  & \textbf{0.5}                  & 0.151          & \textbf{0.207}           & 0.197           & 0.182       & 0.231        & 0.231         & \textbf{0.172}       & 0.194        & 0.187         & 0.411       & 0.535        & 0.531                              \\ \hline
    \end{tabular}
\end{table*}

\begin{table*}[t]
  \centering
  \caption{Performance of IRTest-L with different $\epsilon$ and $f_\mathrm{th}$ across all the three environments}
  \label{table:limited}
  \renewcommand{\arraystretch}{1.2}
  \begin{tabular}{c|c|c|c|ccc|ccc|ccc|ccc}
    \hline
    \multirow{2}{*}{\textbf{Environment}}    & \multirow{2}{*}{\textbf{IS}}  & \multirow{2}{*}{$\bm{\epsilon}$} & \multirow{2}{*}{$\bm{f_{\mathrm{th}}}$} & \multicolumn{3}{c|}{\textbf{AP}}           & \multicolumn{3}{c|}{\textbf{Precision}}    & \multicolumn{3}{c|}{\textbf{Recall}}       & \multicolumn{3}{c}{$\bm{\alpha_2}$}        \\ \cline{5-16} 
                                             &                               &                                  &                                         & \textbf{0k} & \textbf{25k} & \textbf{150k} & \textbf{0k} & \textbf{25k} & \textbf{150k} & \textbf{0k} & \textbf{25k} & \textbf{150k} & \textbf{0k} & \textbf{25k} & \textbf{150k} \\ \hline
\multirow{10}{*}{\textbf{PredatorPrey}}  & \multirow{5}{*}{\textbf{w/o}} & \multirow{3}{*}{\textbf{0.05}}   & \textbf{0.1}                            & 0.129       & 0.145        & 0.144         & 0.0327      & 0.0357       & 0.035         & \textbf{0.982}       & 0.971        & 0.975         & 0.333       & 0.722        & 0.669         \\ \cline{4-4}
                            &                               &                                  & \textbf{0.3}                            & 0.129       & 0.14         & 0.14          & 0.0536      & 0.0581       & 0.0584        & 0.869       & 0.87         & 0.867         & 0.333       & 0.623        & 0.654         \\ \cline{4-4}
                            &                               &                                  & \multirow{3}{*}{\textbf{0.5}}           & 0.129       & 0.136        & 0.137         & \textbf{0.086}       & 0.0919       & 0.0924        & 0.642       & 0.659        & 0.664         & 0.333       & 0.521        & 0.532         \\ \cline{3-3}
                            &                               & \textbf{0.2}                     &                                         & 0.129       & 0.136        & 0.137         & \textbf{0.086}       & \textbf{0.0925}       & \textbf{0.0925}        & 0.642       & 0.662        & 0.667         & 0.333       & 0.541        & 0.542         \\ \cline{3-3}
                            &                               & \textbf{0.5}                     &                                         & 0.129       & 0.145        & 0.146         & \textbf{0.086}       & 0.0923       & 0.0916        & 0.642       & 0.699        & 0.7           & 0.333       & 0.728        & 0.732         \\ \cline{2-16} 
                            & \multirow{5}{*}{\textbf{w}}   & \multirow{3}{*}{\textbf{0.05}}   & \textbf{0.1}                            & 0.129       & 0.145        & 0.144         & 0.0327      & 0.0351       & 0.0348        & \textbf{0.982}       & \textbf{0.976}        & \textbf{0.977}         & 0.333       & 0.667        & 0.633         \\ \cline{4-4}
                            &                               &                                  & \textbf{0.3}                            & 0.129       & 0.145        & 0.15          & 0.0536      & 0.0586       & 0.0593        & 0.869       & 0.868        & 0.863         & 0.333       & 0.697        & 0.812         \\ \cline{4-4}
                            &                               &                                  & \multirow{3}{*}{\textbf{0.5}}           & 0.129       & \textbf{0.151}        & 0.151         & \textbf{0.086}       & 0.0891       & 0.0869        & 0.642       & 0.715        & 0.714         & 0.333       & \textbf{0.84}         & \textbf{1.0}           \\ \cline{3-3}
                            &                               & \textbf{0.2}                     &                                         & 0.129       & 0.149        & \textbf{0.152}         & \textbf{0.086}       & 0.0898       & 0.0873        & 0.642       & 0.709        & 0.714         & 0.333       & 0.804        & 0.969         \\ \cline{3-3}
                            &                               & \textbf{0.5}                     &                                         & 0.129       & 0.145        & 0.142         & \textbf{0.086}       & 0.0911       & 0.0901        & 0.642       & 0.692        & 0.68          & 0.333       & 0.644        & 0.552         \\ \hline
\multirow{10}{*}{\textbf{CoopNavi}}      & \multirow{5}{*}{\textbf{w/o}} & \multirow{3}{*}{\textbf{0.05}}   & \textbf{0.1}                            & 0.468       & 0.526        & 0.566         & 0.153       & 0.158        & 0.168         & \textbf{0.938}       & \textbf{0.942}        & \textbf{0.94}          & 0.333       & 0.481        & 0.602         \\ \cline{4-4}
                            &                               &                                  & \textbf{0.3}                            & 0.468       & 0.519        & 0.536         & 0.25        & 0.307        & 0.316         & 0.849       & 0.864        & 0.863         & 0.333       & 0.477        & 0.523         \\ \cline{4-4}
                            &                               &                                  & \multirow{3}{*}{\textbf{0.5}}           & 0.468       & 0.501        & 0.509         & \textbf{0.408}       & 0.434        & 0.439         & 0.623       & 0.659        & 0.667         & 0.333       & 0.435        & 0.447         \\ \cline{3-3}
                            &                               & \textbf{0.2}                     &                                         & 0.468       & 0.501        & 0.513         & \textbf{0.408}       & 0.434        & 0.441         & 0.623       & 0.659        & 0.672         & 0.333       & 0.434        & 0.455         \\ \cline{3-3}
                            &                               & \textbf{0.5}                     &                                         & 0.468       & 0.511        & 0.537         & \textbf{0.408}       & 0.437        & \textbf{0.459}         & 0.623       & 0.667        & 0.715         & 0.333       & 0.443        & 0.502         \\ \cline{2-16} 
                            & \multirow{5}{*}{\textbf{w}}   & \multirow{3}{*}{\textbf{0.05}}   & \textbf{0.1}                            & 0.468       & 0.429        & 0.481         & 0.153       & 0.15         & 0.153         & \textbf{0.938}       & 0.937        & \textbf{0.94}          & 0.333       & 0.315        & 0.379         \\ \cline{4-4}
                            &                               &                                  & \textbf{0.3}                            & 0.468       & 0.542        & 0.598         & 0.25        & 0.316        & 0.344         & 0.849       & 0.863        & 0.856         & 0.333       & 0.521        & 0.723         \\ \cline{4-4}
                            &                               &                                  & \multirow{3}{*}{\textbf{0.5}}           & 0.468       & \textbf{0.572}        & \textbf{0.658}         & \textbf{0.408}       & 0.448        & 0.403         & 0.623       & 0.78         & 0.804         & 0.333       & \textbf{0.583}        & \textbf{0.944}         \\ \cline{3-3}
                            &                               & \textbf{0.2}                     &                                         & 0.468       & 0.548        & 0.615         & \textbf{0.408}       & \textbf{0.449}        & 0.415         & 0.623       & 0.701        & 0.799         & 0.333       & 0.489        & 0.805         \\ \cline{3-3}
                            &                               & \textbf{0.5}                     &                                         & 0.468       & 0.52         & 0.579         & \textbf{0.408}       & 0.429        & 0.435         & 0.623       & 0.663        & 0.789         & 0.333       & 0.427        & 0.657         \\ \hline
\multirow{10}{*}{\textbf{BipedalWalker}} & \multirow{5}{*}{\textbf{w/o}} & \multirow{3}{*}{\textbf{0.05}}   & \textbf{0.1}                            & 0.195       & 0.213        & 0.219         & 0.100       & 0.102        & 0.104         & \textbf{0.971}       & 0.971        & 0.967         & 0.333       & 0.571        & 0.603         \\ \cline{4-4}
                            &                               &                                  & \textbf{0.3}                            & 0.195       & 0.216        & \textbf{0.225}         & 0.139       & 0.136        & 0.138         & 0.793       & 0.831        & 0.817         & 0.333       & 0.561        & 0.620         \\ \cline{4-4}
                            &                               &                                  & \multirow{3}{*}{\textbf{0.5}}           & 0.195       & 0.218        & 0.222         & \textbf{0.205}       & \textbf{0.208}        & 0.206         & 0.418       & 0.471        & 0.487         & 0.333       & 0.540        & 0.585         \\ \cline{3-3}
                            &                               & \textbf{0.2}                     &                                         & 0.195       & 0.209        & 0.223         & \textbf{0.205}       & 0.202        & \textbf{0.211}         & 0.418       & 0.490        & 0.461         & 0.333       & 0.495        & 0.572         \\ \cline{3-3}
                            &                               & \textbf{0.5}                     &                                         & 0.195       & 0.225        & \textbf{0.225}         & \textbf{0.205}       & \textbf{0.208}        & 0.208         & 0.418       & 0.484        & 0.486         & 0.333       & 0.616        & 0.615         \\ \cline{2-16} 
                            & \multirow{5}{*}{\textbf{w}}   & \multirow{3}{*}{\textbf{0.05}}   & \textbf{0.1}                            & 0.195       & 0.211        & 0.213         & 0.100       & 0.102        & 0.102         & \textbf{0.971}       & \textbf{0.972}        & \textbf{0.972}         & 0.333       & 0.593        & 0.606         \\ \cline{4-4}
                            &                               &                                  & \textbf{0.3}                            & 0.195       & 0.197        & 0.201         & 0.139       & 0.130        & 0.131         & 0.793       & 0.847        & 0.846         & 0.333       & 0.614        & 0.619         \\ \cline{4-4}
                            &                               &                                  & \multirow{3}{*}{\textbf{0.5}}           & 0.195       & \textbf{0.229}        & 0.223         & \textbf{0.205}       & 0.197        & 0.176         & 0.418       & 0.532        & 0.608         & 0.333       & \textbf{0.734}        & \textbf{0.925}         \\ \cline{3-3}
                            &                               & \textbf{0.2}                     &                                         & 0.195       & 0.224        & 0.220         & \textbf{0.205}       & 0.188        & 0.181         & 0.418       & 0.562        & 0.588         & 0.333       & 0.725        & 0.765         \\ \cline{3-3}
                            &                               & \textbf{0.5}                     &                                         & 0.195       & 0.214        & 0.217         & \textbf{0.205}       & 0.184        & 0.182         & 0.418       & 0.572        & 0.584         & 0.333       & 0.666        & 0.726         \\ \hline
\end{tabular}
\end{table*}

\bibliographystyle{IEEEtran}
\bibliography{IEEEabrv,reference.bib}

\begin{IEEEbiography}[{\includegraphics[width=1in,height=1.25in,clip,keepaspectratio]{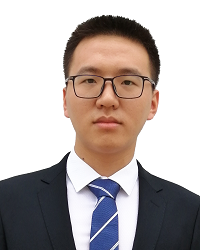}}]{Jingxuan Yang}
  received the bachelor's degree from the School of Mechanical Engineering and Automation, Harbin Institute of Technology, Shenzhen, China, in 2020, and the Ph.D. degree from the Department of Automation, Tsinghua University, Beijing, China, in 2025. His current research interests include intelligent intelligence test, adaptive testing and evaluation, curse of rarity and dense learning.
\end{IEEEbiography}

\begin{IEEEbiography}[{\includegraphics[width=1in,height=1.25in,clip,keepaspectratio]{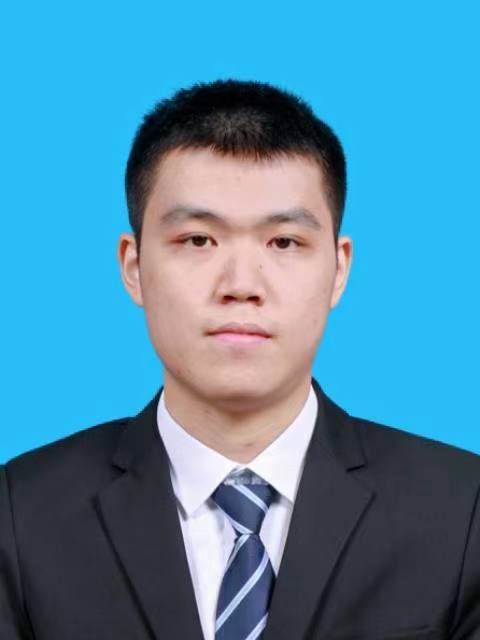}}]{Weichao Xu}
  received the B.S. degree from Department of Automation, Tsinghua University, Beijing, China, in 2023. He is currently working towards the M.S. degree with the Department of Automation, Tsinghua University, Beijing, China. His current research interests include intelligent vehicle-infrastructure cooperative systems, autonomous driving perception and intelligent system testing.
\end{IEEEbiography}

\begin{IEEEbiography}[{\includegraphics[width=1in,height=1.25in,clip,keepaspectratio]{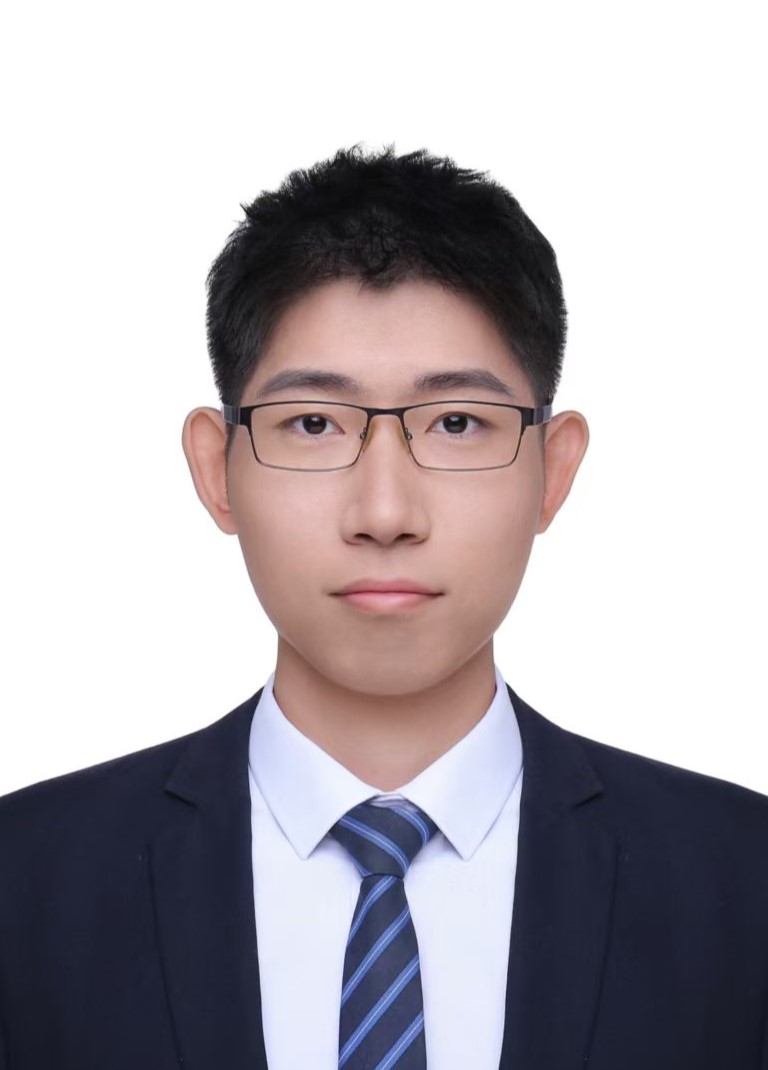}}]{Yuchen Shi}
  received the B.S. degree from Tsinghua University, Beijing, China, in 2021, where he is currently pursuing the Ph.D. degree with the Department of Automation. His research interests include autonomous driving and multi-agent reinforcement learning.
\end{IEEEbiography}

\begin{IEEEbiography}[{\includegraphics[width=1in,height=1.25in,clip,keepaspectratio]{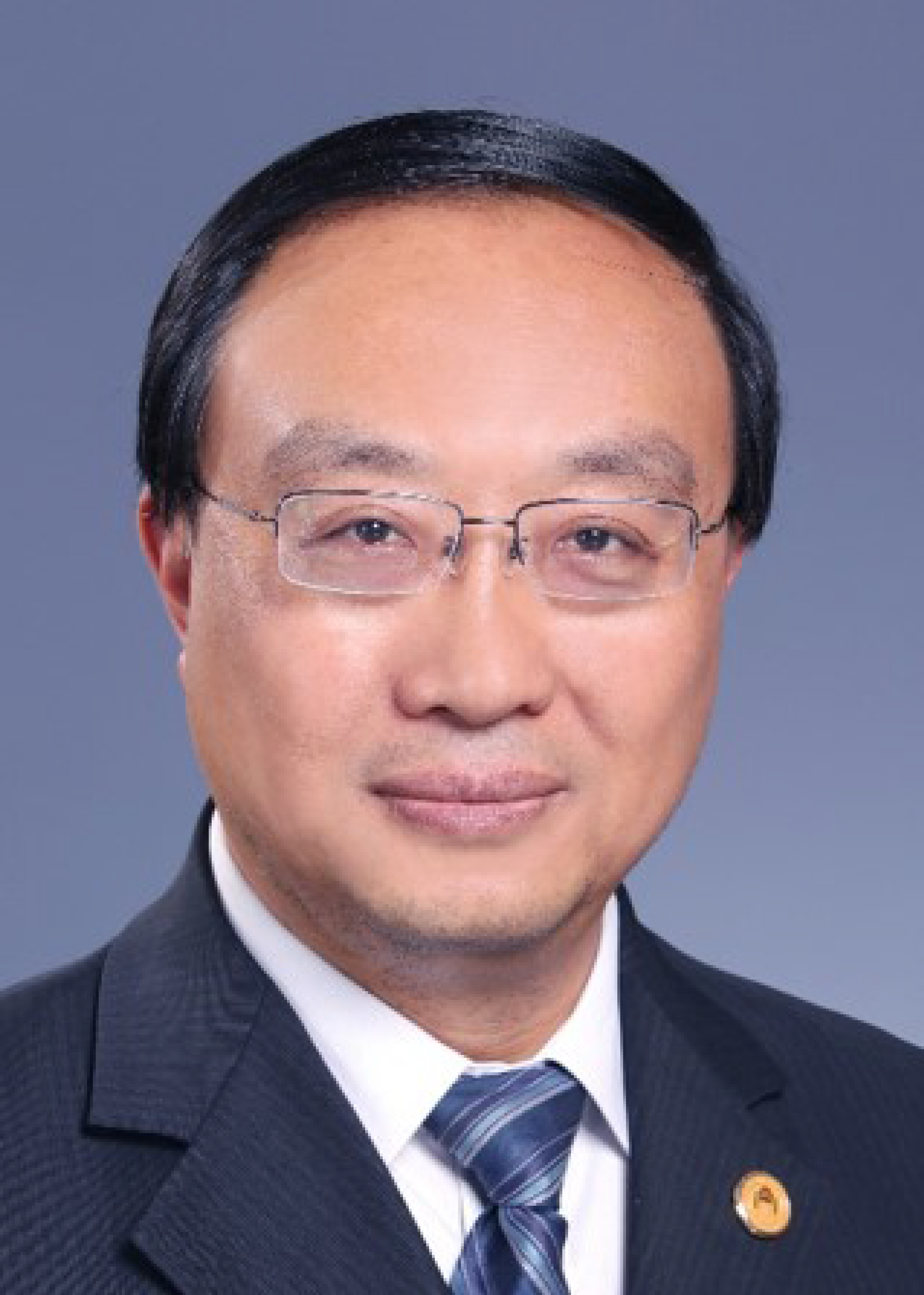}}]{Yi Zhang}
  (Senior Member, IEEE) received the B.S. and M.S. degrees from Tsinghua University, China, in 1986 and 1988, respectively, and the Ph.D. degree from the University of Strathclyde, U.K., in 1995. He is currently a Professor in control science and engineering with Tsinghua University. His current research interests focus on intelligent transportation systems. His active research areas include intelligent vehicle-infrastructure cooperative systems, analysis of urban transportation systems, urban road network management, traffic data fusion and dissemination, and urban traffic control and management. His research fields also cover the advanced control theory and applications, advanced detection and measurement, as well as systems engineering.
\end{IEEEbiography}

\begin{IEEEbiography}[{\includegraphics[width=1in,height=1.25in,clip,keepaspectratio]{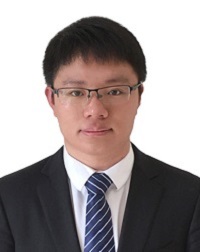}}]{Shuo Feng}
  (Member, IEEE) received the bachelor's and Ph.D. degrees in the Department of Automation at Tsinghua University, China, in 2014 and 2019, respectively. He was a postdoctoral research fellow in the Department of Civil and Environmental Engineering and also an Assistant Research Scientist at the University of Michigan Transportation Research Institute (UMTRI) at the University of Michigan, Ann Arbor. He is currently an Associate Professor in the Department of Automation at Tsinghua University. His research interests lie in the development and validation of safety-critical machine learning, particularly for connected and automated vehicles. He was a recipient of the Best Ph.D. Dissertation Award from the IEEE Intelligent Transportation Systems Society in 2020 and the ITS Best Paper Award from the INFORMS TSL society in 2021. He is an Associate Editor of the \textsc{IEEE Transactions on Intelligent Vehicles} and an Academic Editor of the \textit{Automotive Innovation}.
\end{IEEEbiography}

\begin{IEEEbiography}[{\includegraphics[width=1in,height=1.25in,clip,keepaspectratio]{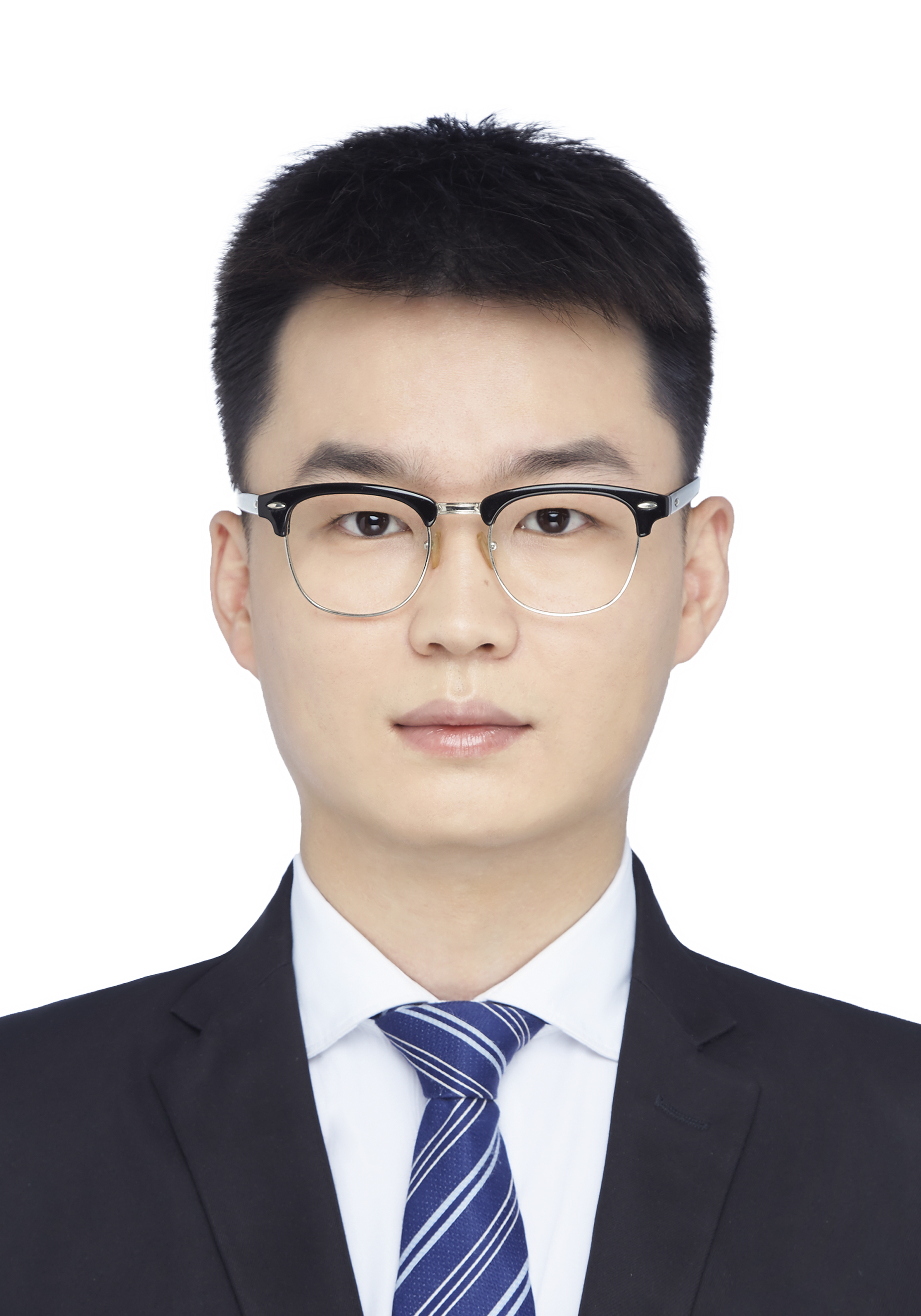}}]{Huaxin Pei}
 received the B.S. degree from University of Electronic Science and Technology of China, China, in 2017, and earned the Ph.D. degree in the control science and engineering with the Department of Automation, Tsinghua University, China, in 2023. He is currently an associate research fellow at QiYuan Lab. His current research interests include intelligent system testing, multi-agent reinforcement learning, autonomous driving, and cooperative driving.
\end{IEEEbiography}

\end{document}